\documentclass[a4paper,11pt]{article}
\pdfoutput=1 % if your are submitting a pdflatex (i.e. if you have
\usepackage{jheppub} % for details on the use of the package, please
\usepackage[T1]{fontenc} % if needed
\usepackage[utf8]{inputenc}

\title{Unity of pomerons from gauge/string duality}

\author[a]{Alfonso Ballon-Bayona,}
\author[b]{Robert Carcassés Quevedo,}
\author[b]{and Miguel S. Costa}

\affiliation[a]{Instituto de Física Teórica, Universidade Estadual
  Paulista,  Rua Dr. Bento Teobaldo Ferraz, 271 - Bloco II,
  01140-070 São Paulo, SP, Brazil}
\affiliation[b]{Centro de Física do Porto e Departamento de Física e Astronomia da Faculdade de Ciências da Universidade do Porto, Rua do Campo Alegre 687, 4169-007 Porto, Portugal}

\emailAdd{aballonb@ift.unesp.br}
\emailAdd{rcarcasses@fc.up.pt}
\emailAdd{miguelc@fc.up.pt}

\abstract{We develop a formalism where the hard and soft pomeron contributions to high energy scattering arise as leading Regge poles of a single kernel in holographic QCD. The kernel is obtained using effective field theory inspired by Regge
theory of a 5-d string theory. It describes the exchange of higher spin fields in the graviton Regge trajectory that are dual to glueball states of twist two. For a specific holographic QCD model we describe Deep Inelastic Scattering in the Regge limit of low Bjorken $x$, finding good agreement with experimental data from HERA. The observed rise of the effective pomeron intercept, as the size of the probe decreases,  is  reproduced by considering the first four pomeron trajectories. In the case of soft probes, relevant to total cross sections, the leading hard pomeron trajectory is suppressed, such that in this kinematical region we reproduce an intercept of 1.09 compatible with the QCD soft pomeron data. In the spectral region of positive Maldelstam variable $t$ the first two pomeron trajectories are consistent with current expectations for the glueball spectrum from lattice simulations.}

\newcommand{\be}{\begin{equation}}
\newcommand{\ee}{\end{equation}}
\newcommand{\bq}{\begin{eqnarray}}
\newcommand{\eq}{\end{eqnarray}}
\newcommand{\bsq}{\begin{subequations}}
\newcommand{\esq}{\end{subequations}}
\newcommand{\bc}{\begin{center}}
\newcommand{\ec}{\end{center}}

\newcommand\lsim{\mathrel{\rlap{\lower4pt\hbox{\hskip1pt$\sim$}} \raise1pt\hbox{$<$}}}
\newcommand\gsim{\mathrel{\rlap{\lower4pt\hbox{\hskip1pt$\sim$}} \raise1pt\hbox{$>$}}}

\begin{document} 

\maketitle
\flushbottom

%%%%%%%%%%%%%%%%%%%%%%%%%%%%%%%%
\section{Introduction}
\label{sec:intro}
%%%%%%%%%%%%%%%%%%%%%%%%%%%%%%%%

Regge theory is the study of the analytic structure of the scattering amplitude in the so called complex angular momentum $J$-plane.
%, which in a pragmatic point of view turns out to be the dual to the Mandelstam variable $s$-plane under a Mellin transform. 
The assumption that the scattering amplitude in the $J$-plane has a pole, the so called pomeron,
such that there are no more singularities at the right of it except at integer values, led to the explanation in the early 60s of the total cross-section behavior with
center of mass energy in $pp$ and $p\bar{p}$ experiments, among others. 
This particular analytic structure suggest that the scattering amplitude in the so called Regge limit of large $s$ at fixed  $t$, is dominated by the interchange of a infinite set of particles of all spins: the ones belonging to the pomeron Regge trajectory \cite{donnachie_pomeron_2002}. 

Regge theory  is particularly appealing since the amplitude obtained in the $s$-plane, $A(s,t) \sim  \Gamma\big(\! - \! j(t) \big) s^{j(t)}$,
analytically continued to the non-physical scattering region of positive $t$, provides a connection with the exchanged spin $J$ bound states of the theory, whose mass
is given by $J=j(M^2)$. This remarkable fact allowed to explain early scattering data when meson trajectories are exchanged. It also led to the proposal that 
there exists another set of resonances with the quantum numbers of the vacuum, associated with the pomeron, which in principle will be a family of the so far unobserved glueballs.
This idea has been supported 
through the years by Donnachie and Landshoff, who showed in the early 90s that Regge theory provides an economical description of  total elastic  cross-sections
\cite{donnachie_total_1992}. This is known as the soft pomeron trajectory, with an intercept of around $j(0)=1.08$, and is well established as a model for total elastic cross-sections of soft particles (for example, $pp$ and $p\bar{p}$ scattering).

%%%%%%%%%%%%%%%
\begin{figure}[t!]
\begin{center}
\includegraphics[height=10cm]{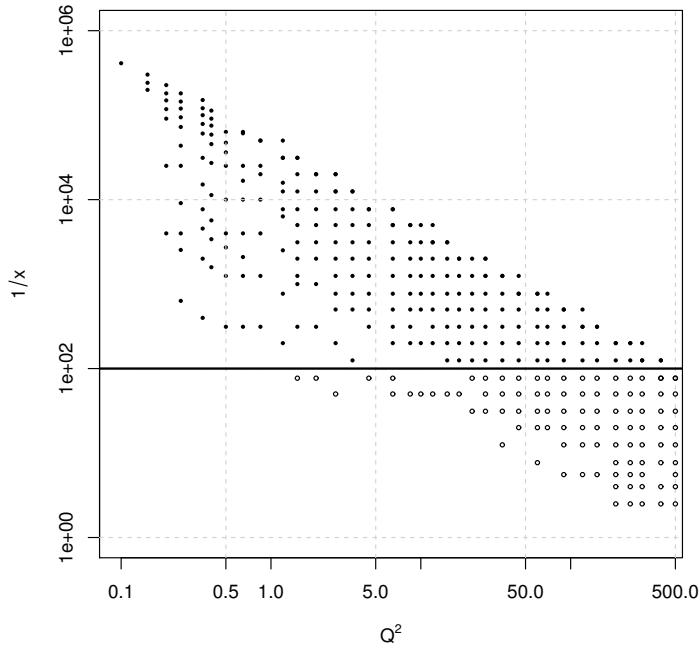}
\end{center}
\vspace{-0.5cm}
\caption{Values of $x$ and $Q^2$ for the data points analysed in this paper \cite{Aaron:2009aa}. Regge kinematics restricts this domain to $x<0.01$.}
\label{fig:HERAdata}
\end{figure}
%%%%%%%%%%%%%%%

\begin{figure}[t!]
\begin{center}
\includegraphics[height=10cm]{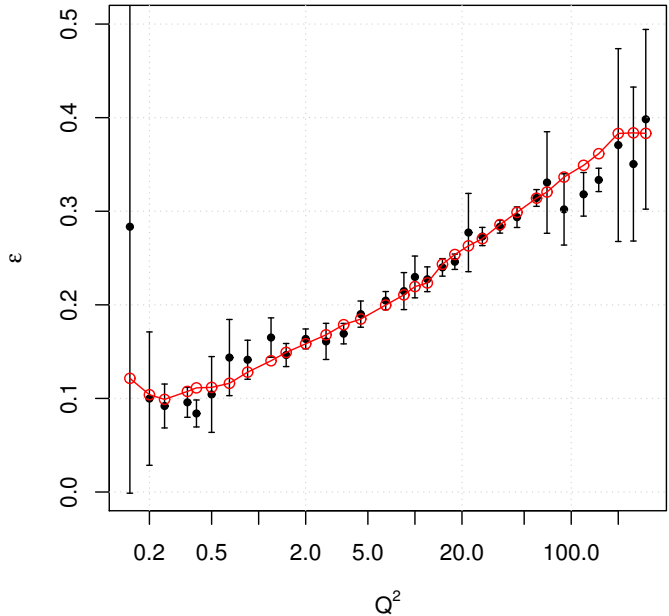}
\end{center}
\vspace{-0.5cm}
\caption{The effective exponent $\epsilon(Q^2)$ in DIS. Black dots are obtained by extrapolating
    the $\log$ of the cross section at fixed $Q^2$ with a straight line in $\log x$. The corresponding error bars are  at $3\sigma$. The red curve is our prediction for the effective exponent using the model proposed in this work.}
\label{fig:intercept}
\end{figure}

Deep inelastic scattering (DIS) is another process where Regge theory is important. In this case we consider
the imaginary part of the amplitude for $\gamma^*p\rightarrow \gamma^*p$, at zero momentum transfer $(t=0)$, which gives the 
total cross section for  the scattering of an off-shell photon with a proton. Single Reggeon exchange then predicts a total cross section
determined by the intercept, $\sigma\sim s^{j(0)-1}$. 
However this story is bit more evolved.
In the $\gamma^*p$ system
there are  two  kinematical quantities: the virtuality of the photon $Q^2$ and the Bjorken $x$, which in the Regge limit is   related to $s$ by $s=Q^2/x$, with $x\ll1$.
When HERA data for DIS scattering came out, it was somehow surprising to observe  that the rise of the cross section with $1/x$ was actually faster than that predicted by the soft pomeron. 
The main difference  is that, instead of using two soft probes for the scattering process, the  off-shell photon virtuality  can be  well above the QCD confining scale. 
What is actually observed is a growth of the intercept with $Q^2$ from about $1.1$ to $1.4$. More concretely, if we write the total cross section as
\begin{equation}
\sigma \big(x,Q^2\big) = f\big(Q^2\big) \,x^{-\epsilon(Q^2)}\,,
\end{equation}
then the exponent $\epsilon$ grows with $Q^2$. Figure \ref{fig:HERAdata} shows the latest data points from HERA experiment, restricted to the region
of low $x$ where Regge kinematics holds. In figure   \ref{fig:intercept}  we see the observed behaviour of the exponent $\epsilon(Q^2)$. 

%%%%%%%%%%%%%%%
%%%%%%%%%%%%%%%

The behaviour of the exponent ${\epsilon(Q^2)}$ for low $Q^2$ is consistent with the observed intercept of the soft pomeron for soft probes, but for hard probes (larger $Q^2$) this is no longer the case, suggesting the existence of another trajectory with a bigger intercept, the so called hard pomeron. The nature of both pomerons, and in particular their relation, remains an unsolved problem
in QCD. Are the soft and hard pomerons the same or distinct trajectories?
Our main motivation in this work is to use holography to shed light into this problem. 

A very interesting proposal to resolve the above puzzle was again put forward by Donnachie and Landshoff \cite{donnachie_small_1998,donnachie_new_2001,donnachie_perturbative_2002,donnachie_elastic_2011,donnachie_pp_2013}.
They proposed that the hard and soft pomerons are distinct trajectories, with the hard pomeron intercept around $1.4$.
The soft pomeron would be dominant in the soft region, since it is already  well established to explain all soft processes,
 and the hard pomeron with a bigger intercept would dominate in the hard processes. More concretely,
 the idea is to write the cross section as
\begin{equation}
\sigma \big(x,Q^2\big) = \sum_n f_n\big(Q^2\big) \, x^{-\epsilon_n}\,,
\label{eq:many_poles}
\end{equation}
where the sum runs over distinct trajectories. Then the effect of summing over several trajectories, which compete with each other as one varies the
virtuality $Q^2$, has the desired effect of producing a varying effective exponent $\epsilon(Q^2)$, as shown in figure \ref{fig:intercept}.
We shall follow this perspective and see that it follows naturally using the gauge/string duality as a tool to study QCD strongly coupled phenomena.

Since QCD is well established as the theory of strong interactions, Regge theory should be encoded on it. 
However,  it turned out to be a remarkable hard problem to deduce the pomeron $J$-plane analytic structure from QCD. 
This fact has its roots in that, even nowadays, we mostly know how to compute QCD quantities perturbatively. The most successful approach has been that of the BFKL pomeron \cite{Fadin:1975cb,Kuraev:1977fs,Balitsky:1978ic}, also known as the hard pomeron,  and its generalisations. 
BFKL, in particular, predicts an amplitude for hadron scattering with a branch cut structure in the complex angular momentum plane. Introducing a momentum cut-off, it yields 
a discrete set of poles that can been confronted with HERA data \cite{kowalski_using_2010}. This approach has two undesired features: it does not cover the non-pertubative region of soft probes and it requires a very large number of poles which results in a very large number of fitting parameters.

\begin{figure}[!b]
    \centering
    \includegraphics[scale=0.9]{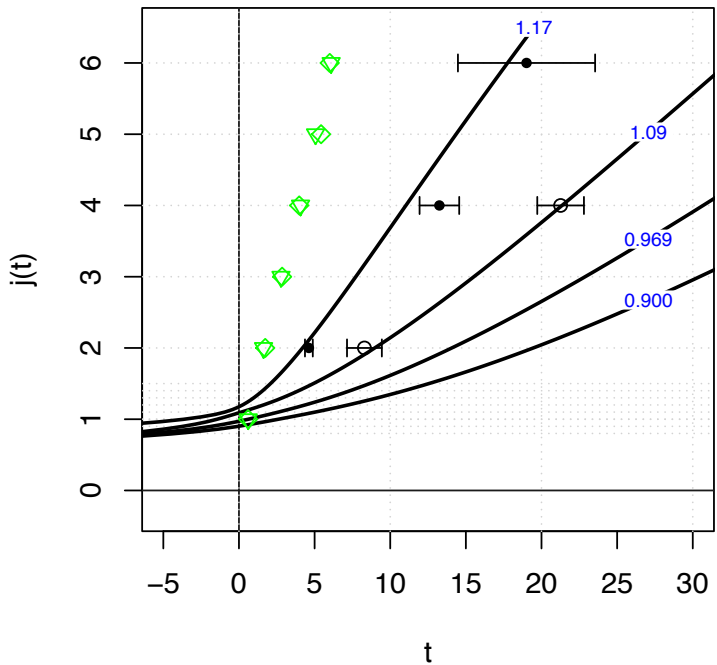}
    \caption{The first four pomeron trajectories found in this paper. The blue labels are the  intercepts of each one. Shown are also the masses  of the higher spin glueballs from lattice QCD data \cite{meyer_glueball_2005,meyer_glueball_2005-1}, which clearly seem to
    belong to the hard and soft pomeron trajectories. In green we plotted the masses of vector mesons, which also contribute to DIS, but are expected to have a lower intercept than the first pomeron trajectories 
    considered in this paper.}
    \label{fig:trajectories}
\end{figure}

The gauge/string duality gives an alternative approach to look at DIS \cite{polchinski_hard_2002,polchinski_deep_2003}. 
Of particular importance to DIS at  low $x$ is the proposal that the the pomeron trajectory is dual to the graviton Regge trajectory \cite{brower_pomeron_2007}.
This work sparked several phenomenological studies that considered QCD processes mediated by pomeron exchange, which in general have been
very successful in reproducing experimental data \cite{hatta_deep_2008,cornalba_saturation_2008,pire_ads/qcd_2008,albacete_dis_2008,hatta_relating_2008,levin_glauber-gribov_2009,brower_saturation_2008,brower_elastic_2009,gao_polarized_2009,hatta_polarized_2009,kovchegov_comparing_2009,avsar_shockwaves_2009,cornalba_deep_2010,dominguez_particle_2010,cornalba_ads_2010,betemps_diffractive_2010,gao_polarized_2010,kovchegov_$r$_2010,levin_inelastic_2010,domokos_pomeron_2009,domokos_setting_2010,brower_string-gauge_2010,costa_deeply_2012,Brower:2012mk, costa_vector_2013,anderson_central_2014, Nally:2017nsp}. 
In this paper we shall explore the above Regge theory ideas for DIS in the  new framework of the gauge/string duality. We  shall test our predictions using the specific 
holographic QCD model proposed in \cite{gursoy_exploring_2008,gursoy_exploring_2008-1,gursoy_improved_2011}. This model 
incorporates  features of the strongly coupled QCD regime, like the spectrum of glueballs and mesons, confinement, chiral symmetry breaking among others.
It is therefore an ideal ground to study processes dominated by the exchange of glueball trajectories. 
Our main findings are summarized in figures \ref{fig:intercept} and \ref{fig:trajectories}. We show that low $x$ DIS data, and in particular the  running of the
effective exponent $\epsilon(Q^2)$,  can be reproduced considering only the first four  pomeron trajectories arising from the graviton trajectory in holographic QCD. The glueball trajectories shown in figure \ref{fig:trajectories} are fixed by DIS scattering data,
but they are also consistent with results of higher spin glueballs from lattice  simulations \cite{meyer_glueball_2005,meyer_glueball_2005-1}. 

 We finish this introduction discussing the validity of our approach. Firstly, our approach is strictly valid in the large-$N_c$ limit where the scattering amplitude
is dominated by planar diagrams. A similar approximation is already done in the BFKL approach to pomeron physics. 
Secondly, and most importantly, in our framework the 't Hooft coupling $\lambda=g_{YM}^2 N_c$ is not restricted to large values.
The reason is that we want to access the regime of hard scattering where the typical QCD coupling is not large. In this regime the stringy corrections to the supergravity action should be important, thus bottom-up approaches to the dual of QCD are more appropriate for our purposes, since they incorporate those corrections in a phenomenological way. For concreteness, we consider the holographic QCD background of \cite{gursoy_exploring_2008,gursoy_exploring_2008-1} and extend it to describe the twist 2 operators associated with higher spin glueballs in the pomeron Regge trajectory. The exchange  of the corresponding dual higher spin fields, that belong to the graviton Regge trajectory, can be resumed
by considering an infinite set  of Witten diagrams and then using standard Regge theory arguments. 
Our work is inspired by the stringy approach of \cite{brower_pomeron_2007}. However, despite its success, the approach in \cite{brower_pomeron_2007} is restricted to large 
$\lambda$ corresponding to the IR regime.
In particular, when describing the equation for higher spin fields we will consider the most general two-derivative terms and take a phenomenological approach 
to fix  the corresponding coefficients. This approach describes the deviations from the conformal limit in the IR region. In the UV region
we impose by hand that the higher spin field equation reproduces free theory, i.e. that they are dual to twist two operators.

This paper is organized as follows. In section \ref{sec:pheno}, we redo the computation by Donnachie and Landshoff  that tries to reproduce DIS data with a hard and a soft 
pomeron, determining the functions $f_n(Q^2)$ in (\ref{eq:many_poles}) from data analysis. Quite remarkably if we translate these functions into the proper gauge/string duality language, they are nothing but wave functions describing the normalizable modes of the graviton Regge trajectory.
Section \ref{sec:gg} presents the necessary formulae to study DIS using the gauge/string duality. The discussion is standard and already scattered in existing literature. In section \ref{seq:pomeron} we focus on the pomeron trajectory, and in particular in constructing the analytic continuation of the spin $J$ equation that describes string fields in the graviton Regge trajectory. This discussion extends that already presented in our previous work \cite{Ballon-Bayona:2015wra}. In section \ref{sec:fit}
we do the data analysis, fitting low $x$ DIS data in the  very large kinematical range of $0.1<Q^2<400$ ${\rm GeV}^2$. Our best fit has a 
$\chi^2$ per degree of freedom of 1.7, without removing presumed outliers existing in data. This leads us to the pomeron Regge trajectories shown in figure \ref{fig:trajectories}. 
We present our conclusions in section \ref{sec:con}.

\section{What is DIS data telling us about holographic QCD?}
\label{sec:pheno}
%%%%%%%%%%%%%%%%%%%%%%%%%%%%%%%%%%

The physics of the  pomeron in the gauge/string duality was uncovered in \cite{brower_pomeron_2007} where pomeron exchange was identified with the exchange of string states in the graviton Regge trajectory. The amplitude for a $2\to2$ scattering process in the Regge limit 
 is then of the general form:
 \begin{equation}
   \label{eq:gen-amplitude}
   A(s,t)=\int dz d\bar{z} \,\phi_1(z)\phi_3(z)\,\mathcal{K}_P(s,t,z,\bar{z})\,\phi_2(\bar{z})\phi_4(\bar{z})\,,
\end{equation}
where the functions $\phi_k(z)$ represent the external scattering waves functions for a given process and 
$\mathcal{K}_P(s,t,z,\bar{z})$ is the so-called kernel of the pomeron which represents the tree level interchange of the aforementioned string states.
 Leaving aside technicalities which will be discussed in section \ref{sec:gg}, the pomeron kernel has the following dual representation
 \begin{equation}
 \mathcal{K}_P(s,t,z,\bar{z}) = \sum_n  f_n \,  s^{j_n(t)-1} \psi _n\big(j_n(t),z\big)\psi _n^*\big(j_n(t),\bar{z}\big)\,,
 \label{eq:Kernel}
 \end{equation}
where the sum runs over the graviton Regge trajectories $j_n(t)$ that arise from quantising string states in the "AdS" box. The quantum number $n$
plays a important role in this work, since  the contribution of the first few pomeron trajectories will be vital to reproduce the DIS cross section. 
The prefactor $f_n$ depends on $j_n(t)$, it factorizes in $z$ and $z'$, and it has  a functional form that depends on the
 specific QCD holographic dual. We shall see that in general it has the form
\begin{equation}
f_n = g\big(j_n(t)\big) \, e^{(1-j_n(t))A(z)}e^{B (z)} \, e^{(1-j_n(t))A(\bar{z})}e^{B (\bar{z})} \,,
\label{eq:prefactor}
\end{equation}
where $A$ is the usual conformal function in the 5D dual metric and the function $B$ will be determined by the background fields, 
for instance by the dilaton field $\Phi$. 
For the specific holographic model used in this paper we will have $B= \Phi-A/2$.

The function $\psi_n(z)$ in (\ref{eq:Kernel}) is the $n$-th excited wave function of a Schr\"{o}dinger problem. We shall see that this fact follows from the spectral representation of the 
propagator of  spin $J$ string fields in the graviton Regge trajectory that are exchanged in the dual $5D$ geometry, analytically continued to $J=j_n(t)$. 
This is a highly non-trivial statement that can  be checked by looking at an amplitude of the form (\ref{eq:gen-amplitude}) and fitting it to 
data. Once the external state functions $\phi_k$  and the specific functional form  (\ref{eq:prefactor}) are fixed, we can use data 
to confirm, or disprove, this fact. More concretely, if we consider a  process dominated in the Regge limit by pomeron exchange and choose a specific holographic QCD
model, we can test this model since the data should {\em know} about the underlying  Schr\"{o}dinger problem formulated in the dual theory.

We consider  DIS, for which the  $p+\gamma^* \rightarrow X$ total cross section  can be computed,  through the optical theorem, from the imaginary part of the amplitude (\ref{eq:gen-amplitude}) for $p+\gamma^* \rightarrow p+\gamma^*$ at zero momentum transfer. 
In this case two of the external state functions, say $\phi_{1,3}(z)$, represent the off-shell photon which couples to the quark bilinear electromagnetic current operator, which is itself
dual to a bulk $U(1)$ gauge field. The insertion of a current operator in a correlation function is then described by a non-normalizable mode of this bulk gauge field. 
The other two functions, $\phi_{2,4}(\bar{z})$, describe  the target proton  in terms of a bulk normalizable mode. We recall that,
in QCD language, the functions $\phi_{k}(z)$ are known as dipole wave functions of the external states.

We wish to find out if the available experimental data is compatible with the holographic recipe, leaving aside technicalities which will be discusses in section \ref{sec:gg}. 
As it is well known, the imaginary part of the amplitude (\ref{eq:gen-amplitude}) at $t=0$ is related to the structure function 
$F_2(Q^2,x)$. Here $Q$ is the offshellness of the spacelike probe photon, whose dependence enters through  the external state wave functions $\phi_{1,3}(z)$. The Maldelstam 
variable $s$ is related to $x$ by the usual expression $s=Q^2/x$, so we are in the low $x$ regime. As a first approximation,
the integration over the variable $z$ in the amplitude (\ref{eq:gen-amplitude})  can be done by considering a Dirac delta function centred at $z\sim 1/Q$. This is  a 
good approximation only for large $Q$, i.e. near the AdS boundary at $z\rightarrow 0$, but it will be enough for the purpose of this section. 
In any case it is a  quick way to gain some insight about the shape of the kernel and the compatibility of our proposal with the experimental data. 
The $\bar{z}$ integral can simply be done because the expression factorizes and the external wave functions  $\phi_{2,4}(\bar{z})$ are normalizable, therefore affecting the  
contribution of each Regge pole by an overall multiplicative constant. 
After these steps the expression for $F_2$, as we will see in the next section, drastically simplifies to
\begin{equation}
    \label{eq:F2_simplified}
    F_2(Q^2,x)= x \sum_n c_n   \left( \frac{Q^2}{x}\right)^{j_n} e^{(-j_n+\frac{1}{2})A(1/Q)}e^{\Phi(1/Q)}\psi_n(1/Q)\,,
\end{equation}
where the $c_n$ do not depent neither on $x$ nor on $Q$, and we denoted by $j_n$ the intercept values of each Reggeon $j_n(t=0)$.
Here we are keeping the right warp factor and dilaton dependence, but if one takes the conformal limit,
$A(z)=-\log(z)$ and  $\Phi=const$, the qualitative result would be the same. Thus we predict a structure function of the form  
\begin{equation}
    \label{eq:F2_DL}
    F_2(Q^2,x)=\sum_n f_n(Q^2)\, x^{1-j_{n}}\,,
\end{equation}
where $f_n(Q^2)$ is the product of known functions and a  Schr\"{o}dinger wave function with quantum number $n$ (the $n$-th excited state).
More concretely, a generic confining potential would produce wavefunctions where its number of nodes can be used to label them: 
the ground state would have one node, the first excited state would have two nodes and so on. 

\begin{figure}[!t]
    \hspace*{-1.6cm}
    \begin{tabular}{ll}
        \includegraphics[scale=0.6]{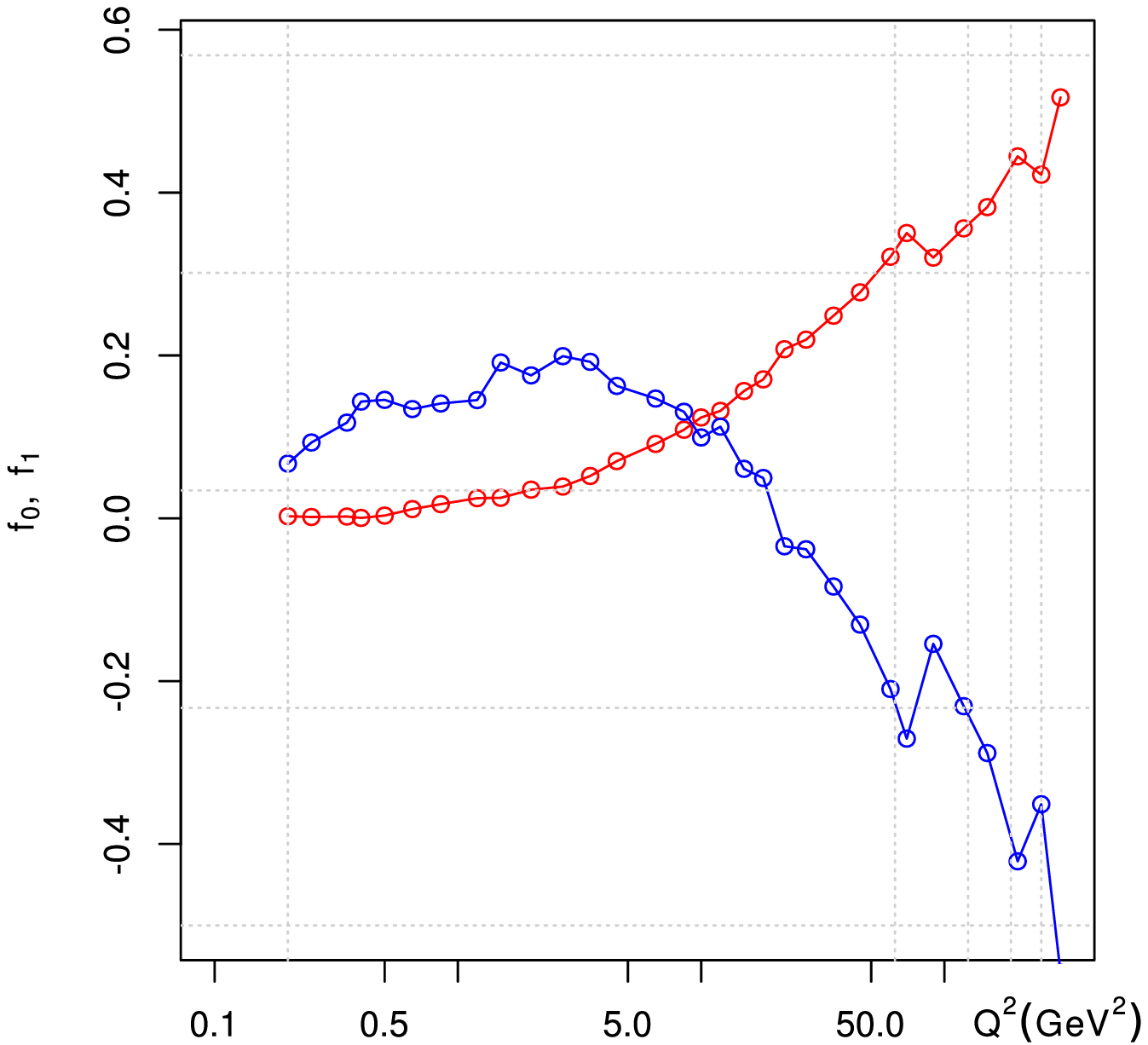}
        &
        \includegraphics[scale=0.6]{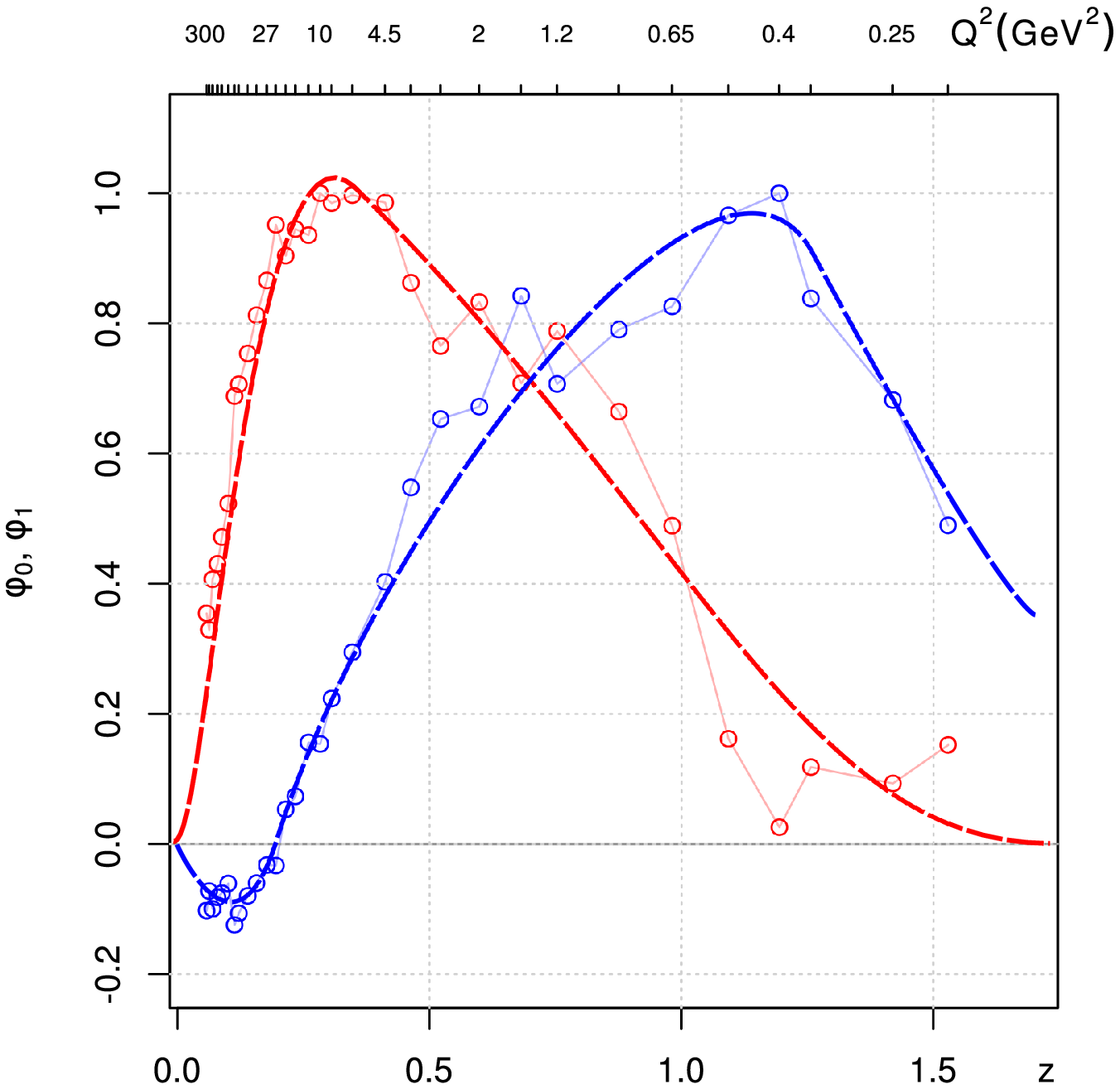}
    \end{tabular}
    \caption{Hard (red)  and soft (blue) pomerons guess from data. Left panel presents the plots of $f_{0,1}(Q)$ similar to \cite{donnachie_small_1998}. The right panel shows the 
    associated wavefunctions $\psi_{0,1}(z)$, after considering the prefactor suggested by the gauge/gravity duality. The values $j_0=1.26$, $j_1=1.08$ have been used, close to what we find later in the paper. Clearly the shape of the wave functions is that of a ground state and of a first excited state of some Schr\"{o}dinger operator.}
    \label{fig:exp_wf}
\end{figure}

Let us now focus on the QCD side of the problem.  Using Regge theory arguments Donnachie and Landshoff  \cite{donnachie_new_2001}
proposed that the  structure function has  precisely the form (\ref{eq:F2_DL}). 
We can do the same reasoning as them. In order to know more about the functions $f_n(Q^2)$ the simplest thing to do is to first consider some fixed values of the $j_n$ that are physically reasonable, like $j_0=1.43$ and $j_1=1.08$. These are reasonable values for the intercepts of the hard and soft pomeron, that are now unified in a single framework, since they appear as distinct Regge trajectories of the dual graviton trajectory in a confining background.
Next, for a fixed value of $Q^2$ we find the best coefficients $f_{0}$ and $f_1$ that match the  data with the formula $f_0\,x^{1-j_0}+f_1\,x^{1-j_1}$, then we can see how these coefficients evolve with $Q^2$. This was already done for a different set of data in \cite{donnachie_new_2001}, which served as a starting point for the authors' proposal for the $f_{0,1}(Q^2)$ functional dependence. Of course the shape of the functions depends on the choice of the intercepts but it is well motivated, given the vast experimental evidence to fix the soft pomeron intercept around $j_1=1.08$. Regarding $j_0$ we should be open to different values, but the expectation is that it will be responsible for the faster growth observer in DIS at higher values of $Q^2$. 
The left panel of figure  \ref{fig:exp_wf} shows the result of this procedure for the values $j_0=1.26$ and $j_1=1.08$, close to what we will show to be the intercepts that give the best fit in our model. The point we want to emphasize is that apparently not much is learned from the shape of these functions.    
    
However, if we divide the functions $f_{0,1}(Q^2)$ by the appropriate functions, as given by (\ref{eq:F2_simplified}),  the putative wave functions $\psi_{0,1}(z=1/Q)$ of the Schr\"{o}dinger problem emerge. This remarkable fact is shown in the right panel of figure \ref{fig:exp_wf}, which clearly meets our expectations.
We should remark that if we use instead $j_1=1.43$, as first suggested by Donnachie and Landshoff, we do not observe the oscillatory behavior expected for $\psi_1$, suggesting perhaps that this value is unphysical. In fact it is known that recent data suggests a smaller $j_0$ \cite{donnachie_elastic_2011}. Indeed, as soon as we get below certain threshold value for $j_1$ the oscillatory behaviour becomes evident with the first node of $\psi_1$ localized very close to the boundary. Moreover, the form of the wavefunctions in our kernel will be very similar to the dashed lines in the figure. We take this as a strong evidence that the DIS data has encoded the dynamics suggested by holographic QCD.  In the next sections we will proceed to phenomenologically construct the effective Schr\"{o}dinger potential that leads to the wavefunctions $\psi_n(z)$ that fit best the data.

The discussion of this section was oversimplified, but it brings out the main idea. In practice, the integral over $z$ in the dual representation of the amplitude (\ref{eq:gen-amplitude}) is not localized, since we also consider  lower values of $Q^2$. Also, to get a reasonable fit to the data we need to include the first four pomeron Regge trajectories. This is fine because those trajectories will be dominant with respect to the $1/s$ corrections to the leading hard pomeron trajectory. Eventually one would also need to include the 
exchange of meson Regge trajectories, but that is for now left out of our work, since those trajectories are still suppressed with respect to the first four Pomerons.

%%%%%%%%%%%%%%%%%%%%%%%%%%%%%%%%%%
\section{Low $x$ DIS  in holographic QCD}
\label{sec:gg}
%%%%%%%%%%%%%%%%%%%%%%%%%%%%%%%%%%
In this section we present the essential ingredients of the effective field theory description for low $x$ DIS in holographic QCD \footnote{Holographic approaches for DIS in the large $x$ regime can be found in \cite{polchinski_deep_2003,BallonBayona:2007qr,BallonBayona:2008zi,BallonBayona:2010ae,Koile:2013hba,Koile:2015qsa}.}. First we briefly describe the kinematics of DIS and its connection to forward Compton scattering amplitude via the optical theorem. Then we present the holographic description of that amplitude, in the Regge limit, via the exchange of higher spin fields. We finish the section deriving a formula similar to (\ref{eq:F2_DL}) which encodes the Regge pole contribution to  the DIS structure functions. 
%%%%%%%%%%%%%%%%%%%%%%%%%%%%%%%%%%
\subsection{Kinematics}
%%%%%%%%%%%%%%%%%%%%%%%%%%%%%%%%%%
In DIS a beam of leptons scatters off a hadronic target of momentum $P$. Each lepton interacts with the hadron through the exchange of a virtual photon of momentum $q$. DIS is an inclusive process because the scattering amplitude involves the sum over all possible final states.  The relevant quantities  are the virtuality $Q^2=q_{\mu} q^{\mu}$ and the Bjorken variable $x=-Q^2/(2 P \cdot q)$.  Another quantity is the Mandelstam variable $s = -(P + q)^2$, describing the squared center of mass energy of the virtual photon-hadron scattering process. 

The DIS  cross section is described in terms of the hadronic tensor 
\begin{equation}
W^{\mu \nu} = i \int d^4 x \, e^{i q \cdot x} \langle H , P \vert \Big [ J^{\mu} (x) , J^{\nu}(0) \Big ] \vert  H, P \rangle  \,,
\end{equation}
where $J^{\mu}$ is the electromagnetic current operator and $\vert H , P \rangle$ denotes a hadronic state $H$ of momentum $P^\mu$. Current conservation and Lorentz invariance
imply that $W^{\mu \nu}$ has the decomposition
\begin{equation}
W^{\mu \nu} = 
F_1\big(x,Q^2\big) \left ( \eta^{\mu \nu} - \frac{q^{\mu} q^{\nu}}{Q^2} \right ) +  \frac{2 x}{Q^2}  F_2\big(x,Q^2\big) \left (P^{\mu} + \frac{q^{\mu}}{2x} \right ) \left (P^{\nu} + \frac{q^{\nu}}{2x} \right )  . 
\end{equation}
The Lorentz invariant quantities $F_1(x,Q^2)$ and $F_2(x,Q^2)$ are the structure functions of DIS. They determine completely the DIS  cross section and provide information regarding the partonic distribution in hadrons.  

The optical theorem relates the hadronic tensor $W^{\mu \nu}$ to the imaginary part of the scattering amplitude describing forward Compton scattering.
For a  photon of incoming momentum $k_1$ and outgoing momenta $-k_3$, and for a hadron of incoming momentum $k_2$ and outgoing momenta $-k_4$,
this amplitude  admits the following decomposition 
\begin{equation}
A^{\text{FC}}(q,P)= 
i (2 \pi)^4 \delta^4 \Big(\sum_i k_i\Big) \Big\{ \xi _T^2(q) \, \tilde{F}_1\big(x,Q^2\big)+\frac{2x}{Q^2}\left(\xi^T(q)\cdot P\right)^2\tilde{F}_2\big(x,Q^2\big) \Big\} \, ,     
\label{AFC}
\end{equation}
where we identified  $q = k_1 = - k_3$ and $P=k_2=-k_4$, and defined the transverse projection of the virtual photon polarization $\xi^{\mu}$ as 
\begin{equation}
\xi_{\mu }^T(q) = \left ( \eta_{\mu \nu} - \frac{q_\mu q_\nu}{Q^2} \right ) \xi^\nu \,. 
\end{equation}
The DIS structure functions are then extracted from the relations
\begin{equation}
F_1 (x,Q^2) = 2 \pi {\rm Im} \tilde F_1 (x,Q^2) \, , \qquad\qquad
F_2 (x,Q^2) = 2 \pi {\rm Im} \tilde F_2 (x,Q^2) \, .
\label{StructureFunctions}
\end{equation}

In DIS there are two interesting limits that are usually considered. The first is the Bjorken limit, where $Q^2 \to \infty$ with $x$ fixed. In this limit perturbative QCD provides a good description of the experimental data in terms of partonic distribution functions. 
%\footnote{DIS experiments at the Stanford Linear Accelerator Center (SLAC) in the late 60's were important to establish QCD as the theory of hadrons} 
The second interesting case is the limit of $s \to \infty$, the so-called Regge limit of DIS, for which $Q^2$ is fixed and $x\approx-Q^2/s$ is very small.
In this limit the hadron becomes a dense gluon medium so that the picture of the hadron made of weakly interacting partons is no longer valid.
As explained in section \ref{sec:pheno}, in this paper we investigate DIS in the Regge limit (low $x$) from the perspective of the pomeron in holographic QCD, which encodes the dynamics of the  dense gluon medium. 
We develop a five dimensional model for the graviton Regge trajectory for a family of backgrounds dual to QCD-like theories in the large-$N$ limit. 
Our formalism leads to the existence of a set of leading Regge poles describing DIS in the Regge limit, the first two interpreted as the hard and soft pomerons.

%%%%%%%%%%%%%%%%%%%%%%%%%%%%%%%%%%
\subsection{Regge theory  in holographic QCD}
\label{ReggeTheory}
%%%%%%%%%%%%%%%%%%%%%%%%%%%%%%%%%%

Let us now consider the computation of the  forward Compton scattering amplitude in holographic QCD. We are interested in
elastic scattering between a virtual photon and a scalar particle with incoming momenta $k_1$ and $k_2$, respectively.
In light-cone coordinates $(+,-,\perp)$, for the  external off-shell photon with virtuality $Q^2$ we take 
\begin{equation}
k_1 = \left(\sqrt{s} , - \frac{Q^2}{\sqrt{s}},0\right) \,,\qquad 
-k_3  =  \left(\sqrt{s} , \frac{ q_\perp^2 - Q^2}{\sqrt{s}} , q_{\perp}\right),
\label{eq:momenta_photon}
\end{equation}
while for the target hadron of mass  $M$ we take 
\begin{equation}
k_2 =  \left(\frac{M^2}{\sqrt{s}} , \sqrt{s} , 0 \right) \, , \qquad 
- k_4 = \left(\frac{M^2 + q_{\perp}^2}{\sqrt{s}} , \sqrt{s} , - q_{\perp}\right) . 
\label{eq:momenta_proton}
\end{equation}
The Regge limit corresponds to $s \gg t = - q_{\perp}^2$ and the case $t=0$ corresponds to forward Compton scattering. The momenta 
$k_1$ and $k_2$ are, respectively, identified with the $q$ and $P$ defined in the previous subsection.
As explained, we will extract  the DIS structure functions from the forward Compton amplitude.

First we define with generality the holographic model that may be used. We need to define the external states in DIS and the 
 interaction between them that is dominated by a t-channel exchange of higher spin fields (those in the graviton Regge trajectory). 
Later on, to compare with the data, we will use a specific holographic QCD model \cite{gursoy_exploring_2008,gursoy_exploring_2008-1}, 
but for now we will write general formulae that can be used in other models. 

The string dual of QCD will have a dilaton field and a five-dimensional metric that are, respectively, dual to the Lagrangian and the energy-momentum tensor.
In the vacuum those fields will be of the form
\begin{equation}
ds^2 = e^{2 A(z)} \left[ dz^2 + \eta_{\mu \nu} dx^\mu dx^\nu \right]    \,, \qquad
\Phi = \Phi(z) \,, \label{IHQCDBackground}
\end{equation}
for some functions $A(z)$ and $\Phi(z)$ whose specific form we assume is known. 
We shall use greek indices in the boundary, with flat metric $ \eta_{\mu \nu}$.
We are defining the warp factor $A(z)$ with respect to the string frame metric.

In DIS the external photon is a source for the conserved $U(1)$ current $\bar{\psi} \gamma^\mu \psi$,
where the quark field $\psi$ is associated to the open string sector. 
The five dimensional dual of this current is a massless $U(1)$ gauge field $A$. We shall assume that this 
field is made out of open strings and that is minimally coupled to the metric, so  its effective action has the 
following simple form
\begin{equation}
S_A=-\frac{1}{4}\int d^5 X \sqrt{-g} \, e^{-\Phi }F^{ab}F_{ab}\,,   
\label{Maxwell}
\end{equation}
where $F=dA$ and we use the notation $X^a=(z,x^\alpha)$ for five-dimensional points. We could in principle have higher order terms in $F$ and other couplings to the metric 
in the action, but for the sake of simplicity we shall work with this action.
As reviewed in appendix \ref{U(1)field}, 
after a convenient gauge choice, the gauge field components describing a boundary plane wave solution with polarization $\xi$  take the form
\begin{equation}
A_{\mu }(x,z)=\xi _{\mu} \, e^{i q \cdot x}f\!\left(Q^2,z\right)  , \qquad \qquad 
A_z(x,z)=e^{iq \cdot x}g\!\left(Q^2,z\right)  ,
\label{eq:ansatzA}
\end{equation}
where $f$ and $g$ satisfy the equations
\begin{equation}
e^{\Phi -A}\partial _z\left(e^{A-\Phi }\partial _zf\right) - Q^2f=0 \, , \qquad \qquad 
g=-i\frac{q\cdot \xi }{Q^2} \, \partial_z f \,. 
\label{eq:eqsA}
\end{equation}
Since we are computing an amplitude with a source for the electromagnetic current operator  $\bar{\psi} \gamma^\mu \psi$,  
the boundary conditions for $f$ are those of a non-normalizable mode, i.e.  $f\left(Q^2,z=0\right)=1$ and $f\left(Q^2,z\to \infty \right)=0$. 
Note that the field strength also takes a plane wave form $F_{ab}(x,z) = e^{iq\cdot x} F_{ab }(q,z)$. As we shall see, a useful quantity is the stress-like tensor 
\begin{eqnarray}
F_{\mu a}(q,z)F_{\ \nu }^{a}(-q,z) = -e^{-2A(z)}\Big \{ \xi _{\mu }^T(q) \,\xi_{\nu }^T(q)\Big[Q^2f^2+\left(\partial _zf\right){}^2\Big] + q_{\mu }q_{\nu }\,\xi_T^2(q)f^2
\Big \}. \label{FF}
\end{eqnarray}

For the target we consider a scalar field $\Upsilon$ that represents an unpolarised proton. This hadronic state is described by a normalizable mode of the form
\begin{equation}
\Upsilon(x,z)= e^{i P\cdot x} \upsilon(z)\,.
\end{equation}
The specific details will not be important. We will simply assume that we can make the integration over
the point where this field interacts with  the higher spin fields. The effect of such  
an overall factor can be absorbed in the coupling constant.

The next step in our construction is to introduce the higher spin fields $h_{b_1 \dots b_J}$ that will mediate the 
 interaction terms between the external fields of the scattering process. 
These fields are dual to the spin $J$ twist two operators made of the gluon field
that are in the leading Regge trajectory. 
There are also other twist two operators made out of the quark bilinear. However, as we shall see, the corresponding Regge trajectories are subleading with respect to the first pomeron trajectories here 
considered. Noting that the higher spin field is in the closed string sector, and that the external fields are in the open sector,  we shall  consider the minimal coupling 
\begin{eqnarray}
\kappa_J\int d^5 X \sqrt{-g} \, e^{-\Phi } \left ( F_{b_1 a} D_{b_2} \dots D_{b_{J-1}}F_{\, \, \, b_J}^a \right ) \, h^{b_1\dots b_J} \, , 
\label{eq:couplingA}
\end{eqnarray}
for the gauge field $A_a$ and  
\begin{eqnarray}
\bar{\kappa}_J \int d^5 X \sqrt{-g} \, e^{-\Phi } \, \left ( \Upsilon D_{b_1} \dots D_{b_J} \Upsilon \right ) \,  h^{b_1 \dots b_J} \, ,
\label{eq:couplingScalar}
\end{eqnarray}
for the scalar field $\Upsilon$. The higher spin field $h_{b_1 \dots b_J}$ is totally symmetric, traceless and satisfies the transversality condition $\nabla^{b_1} h_{b_1 \dots b_J}=0$. The latter fact
implies we do not need to worry in which external fields the derivatives in (\ref{eq:couplingA})  and (\ref{eq:couplingScalar}) act. However, there can be other couplings to the derivatives of the dilaton field 
and also to the curvature tensor. Here we consider only this leading term in a strong coupling expansion (that is, the first term in the derivative expansion of the effective action).
Below we simply assume that the higher spin field has a propagator, without specifying its form. In the next section we focus on the dynamics of this field in detail.

In the Regge limit, the amplitude describing the spin J exchange between the incoming gauge field $A_a^{(1)} \sim e^{i k_1 \cdot x}$ and scalar field $\Upsilon^{(2)} \sim e^{i k_2 \cdot x}$ can be written as 
\begin{align}
A_J\left(k_1,k_2,k_3,k_4\right)=
\left(i\kappa_J\right)\left(i\bar{\kappa}_J\right)\int  d^5 X \int d^5 \bar{X} \sqrt{-g(z)} \, e^{-\Phi(z)} \sqrt{-g(\bar{z})} \, e^{-\Phi (\bar{z})} \cr 
\times  \left ( F_{- \, a}{}^{(1)}(X) \partial _-^{J-2}F_{\, \, \,  -}^a{}^{(3)}(X) 
\right )\, \Pi ^{-\dots-,+\dots +}(X,\bar{X}) \left ( \Upsilon ^{(2)}(\bar{X}) \bar{\partial} _+^J\Upsilon ^{(4)}(\bar{X}) \right ) .
\label{eq:witten_diagram}
\end{align}
The fields $A_a^{(3)} \sim e^{i k_3 \cdot x}$ and $\Upsilon^{(4)} \sim e^{i k_4 \cdot x}$ represent the outgoing gauge and scalar fields. 
The tensor $\Pi^{a_1 \dots a_J, b_1 \dots b_J}(X,\bar{X})$ represents the propagator of the spin J field.  After some algebra the amplitude takes the form 
\begin{align}
A_J \left(k_1,k_2,k_3,k_4\right)= i V \left(\frac{4}{s}\right)  \frac{\kappa_J\bar{\kappa}_J}{2^{J}} \int dz\int d\bar{z}\sqrt{g_{3}(z)}e^{-\Phi (z)}\sqrt{g_{3}(\bar{z})}e^{-\Phi (\bar{z})} \cr 
\times F_{- \, a}\left(k_1,z\right)F_{\, \, \, -}^a\left(k_3,z\right) \Upsilon \left(k_2,\bar{z}\right) \Upsilon \left(k_4,\bar{z}\right)[S(z,\bar{z})]^J 
 \left[e^{A(z)}e^{A(\bar{z})}G_J(z,\bar{z},t)\right]  ,  
\end{align}
where
 $V = (2\pi )^4 \delta ^4 (\sum k_i)$ and $g_{3}(z)$ is the determinant of the  3-d transverse metric given by $ds_{3}^2 = e^{2A(z)} \big[ dz^2 + dx_\perp^2 \big]$.
This is the metric on the  transverse space of the dual scattering process. The local energy squared for the  dual scattering process is given by 
 $S(z,\bar{z})=s \,  e^{-A(z)}e^{-A(\bar{z})}$.
The function $G_J(z,\bar{z},t) = \int d^2l_\perp e^{-iq_\perp\cdot l_\perp} G_J(z,\bar{z},l_\perp)$  is the Fourier transform of the integrated propagator
for a field of even spin $J$,
\begin{eqnarray}
\hspace{-1cm}G_J(z,\bar{z},l_\perp) = i \,2^{J} \left [ e^{A(z)} e^{ A(\bar{z})} \right ]^{1-J} 
 \int \frac{ d w^{+} d w^{-}}{2} \Pi_{+ \dots +,- \dots -} (w^{+},w^{-},\ell_{\perp},z,\bar{z})  \, ,    
\label{eq:transverse_prop} 
\end{eqnarray}
and the light-cone coordinates $w^{\pm}$ are defined by the relation $x-\bar{x}=(w^{+},w^{-},l_\perp)$ with $l_\perp = x_\perp - \bar{x}_\perp$. 
For the case of forward Compton scattering we have that $k_1=-k_3=q$, $k_2=-k_4=P$ and $t=0$. Summing over the contribution of the fields with spin $J=2,4,\dots$, and using the result in (\ref{FF}), 
we obtain the amplitude 
\begin{align}
A^{\text{FC}}(q,P)& = i  V \left(4 Q^2 \right)
\int dz\int d\bar{z}\,P_{24}\left(P^2,z'\right) 
\label{AFCv2}
\\
&\times \left[\frac{s}{2Q^2} \, \xi _T^2(q)P_{13}{}^{(1)} \!\left(Q^2,z\right) +\frac{2}{s}\left(\xi ^T(q)\cdot P\right)^2P_{13}{}^{(2)}\!\left(Q^2,z\right)\right]
\, \chi (s,t=0,z,z')\,,
\nonumber
\end{align}
where we have defined 
\begin{eqnarray}
P_{13}{}^{(1)}\!\left(Q^2,z\right)&=&\sqrt{g_{3}(z)} \, e^{-\Phi (z)}e^{-2A(z)}f^2 \, , \cr 
P_{13}{}^{(2)}\!\left(Q^2,z\right)&=&\sqrt{g_{3}(z)} \, e^{-\Phi (z)}e^{-2A(z)}\left[f^2+\frac{1}{Q^2}\left(\partial _z f\right){}^2\right]  , 
\label{eq:P13}
\\
P_{24}\!\left(P^2,\bar{z}\right)&=&\sqrt{g_{3}(\bar{z})} \, e^{-\Phi (\bar{z})} \,  \Upsilon^2\left(P^2,\bar{z}\right) \, , 
\nonumber
\end{eqnarray}
and  $\chi (s,t,z,\bar{z})$ is the eikonal phase defined by
\begin{equation}
\chi (s,t,z,\bar{z})= 
%\left(\frac{1}{2s}\right)\sum_{J=(2,4,\dots)} \frac{\bar{\kappa}_J \kappa_J}{(-2)^J}  %[\hat{s}(z,z')]^J\left[e^{A(z)}e^{A(z')} G_J(z,z',t)\right] \cr &=&
-\left(\frac{\pi }{4s}\right)\int \frac{d \, J}{2\pi i}\frac{[S(z,\bar{z})]^J+[-S(z,\bar{z})]^J}{\sin(\pi J)}\frac{ \kappa_J\bar{\kappa}_J}{2^J}  \, e^{A(z)+A(\bar{z})}G_J(z,\bar{z},t) \,.
\label{EikonalPhase}
\end{equation}
In (\ref{EikonalPhase}) we  used a Sommerfeld-Watson transform to convert the sum in $J=2,4,\dots$  into an integral in the complex J-plane. 
Comparing the expressions (\ref{AFC}) and (\ref{AFCv2}) for the forward Compton amplitude, and using (\ref{StructureFunctions}), we extract the DIS structure functions for
holographic QCD:
\begin{eqnarray}
\!\!\!\!\!\!\!\!
2xF_1\!\left(x,Q^2\right)\!&=& 2\pi \left(4 Q^2\right)\! \int \!dz\!\int\! d\bar{z}\,P_{13}{}^{(1)}\!\left(Q^2,z\right)P_{24}\!\left(P^2,\bar{z}\right)\text{Im}\big[\chi (s,t=0,z,\bar{z})\big],
\nonumber\\
F_2\!\left(x,Q^2\right)\!&=& 2\pi \left(4 Q^2\right) \!\int\! dz\!\int \!d\bar{z}\,P_{13}{}^{(2)}\!\left(Q^2,z\right)P_{24}\!\left(P^2,\bar{z}\right)\text{Im}\big[\chi (s,t=0,z,\bar{z})\big].
\end{eqnarray}

%%%%%%%%%%%%%%%%%%%%%%%%%%%%%%%%%%
\subsection{Regge poles}
%%%%%%%%%%%%%%%%%%%%%%%%%%%%%%%%%%
\label{subsec:Reggepoles}

In the next section we will describe the dynamics of a higher spin field $h_{a_1 \dots a_J}$. In particular, we shall see how 
the propagator  $G_J(z,z',t)$ admits a spectral representation associated to a Schr\"{o}dinger problem that describes massive spin $J$ glueballs. 
Assuming that such Schr\"{o}dinger potential admits an infinite set of bound states for fixed $J$, we will show that
\begin{eqnarray}
G_J(z,\bar{z},t) &=& e^{B(z)+B(\bar{z})}  \sum_n\frac{\psi _n(J,z) \,\psi _n^*(J,\bar{z})}{t_n(J)-t}  \, . \label{SpectralDecomposition}
\end{eqnarray}
The function $B(z)$ depends on the particular holographic QCD model and will be obtained below for backgrounds of the form (\ref{IHQCDBackground}). The eigenfunctions and eigenvalues of the Schr\"{o}dinger equation are $\psi _n(J,z)$ and $t_n(J)$, respectively. Plugging this result in (\ref{EikonalPhase}) and deforming the contour integral, so that we pick up the contribution from the Regge poles $j_n(t)$, we find that \footnote{This procedure is standard in Regge Theory (see e.g. \cite{donnachie_pomeron_2002}).}
\begin{align}
&\chi (s,t,z,\bar{z})=-\left(\frac{\pi }{4s}\right)
 e^{A(z)+ A(\bar{z})+ B (z)+ B (\bar{z})}
\\
&\times
\sum_n \frac{\kappa _{j_n(t)}\bar{\kappa}_{j_n(t)}}{2^{j_n(t)}} 
\left [ \cot\!\left(\frac{\pi}{2} j_n(t)\right) + i \right ] [S(z,\bar{z})]^{j_n(t)}  
 j_n'(t) \,  \psi _n\big(j_n(t),z\big) \,\psi _n^*\big(j_n(t),\bar{z}\big) \,.
\nonumber
\end{align}
In DIS this result implies that the structure functions $F_1\!\left(x,Q^2\right)$ and $F_2\!\left(x,Q^2\right)$ take the Regge form 
\begin{eqnarray}
2x F_1\!\left(x,Q^2\right)&=& \sum_n g_n     \,x^{1-j_n(0)} Q^{2j_n(0)} \bar P_{13}{}^{(1,n)}\big(Q^2\big)  \,, \cr 
F_2\!\left(x,Q^2\right) &=& \sum_n g_n  \,x^{1-j_n(0)} Q^{2j_n(0)}\bar P_{13}{}^{(2,n)}\big(Q^2\big) \,, 
\label{F1F2Regge}
\end{eqnarray}
where we have defined the functions 
\begin{equation}
\bar{P}_{13}{}^{(i,n)}\!\left(Q^2\right)=\int dz \, P_{13}{}^{(i)}\!\left(Q^2,z\right)   e^{\left(1 - j_n(0) \right)A(z)}e^{B (z)}\psi _n\big(j_n(0),z\big) \,,\quad (i=1,2) \, ,
%\bar{P}_{24}{}^{(n)}\!\left(P^2\right)&=&\int dz \, P_{24}\!\left(P^2,z\right)\left[e^{\left(1 - j_n(0) \right) A(z)}e^{-B (z)}\psi _n{}^*\!\left(j_n(0),z\right)\right]  .
\label{eq:P13bar}
\end{equation}
and the couplings 
\begin{equation}
g_n = - 2\pi ^2 \frac{\kappa _{j_n(0)}\bar{\kappa}_{j_n(0)}}{2^{j_n(0)}}\, j_n'(0)\int dz \, P_{24}\!\left(P^2,z\right) e^{\left(1 - j_n(0) \right) A(z)}e^{B (z)}\psi^* _n\big(j_n(0),z\big) \,.
\label{eq:couplings}
\end{equation}
Notice that in (\ref{F1F2Regge}) we have already used the relation $s = Q^2/x$, valid in the Regge limit of DIS. The couplings $g_n$ include our ignorance of the 
hadron dual wave function, which appears in the integrand of (\ref{eq:couplings}), as well as the local couplings in the dual picture between the external fields and the spin $J$
field. The formula (\ref{F1F2Regge}) has the expected form  (\ref{eq:F2_DL}) advocated by Donnachie and Landshoff.

%%%%%%%%%%%%%%%%%%%%%%%%%%%%%%%%%%
\section{Pomeron in holographic QCD}
\label{seq:pomeron}
%%%%%%%%%%%%%%%%%%%%%%%%%%%%%%%%%%

In the large $s$ scattering regime the lowest twist two operators dominate in the OPE of the currents appearing in the computation of the hadronic tensor. 
Therefore we consider here the interchange of the gluonic $\mathcal{O}_J$ twist 2 operators of the form 
\begin{equation}
{\cal O}_J \sim {\rm tr} \left[  F_{\beta\alpha_1} D_{\alpha_2} \dots  D_{\alpha_{J-1}} F^{\beta}_{\ \alpha_J}  \right] ,
\label{eq:spinJOperator}
\end{equation}
where $D$ is the QCD covariant derivative. In the singlet sector there are also 
twist 2 quark operators of the form $\bar{\psi} \gamma_{\alpha_1} D_{\alpha_2} \cdots D_{\alpha_J} \psi$,  
but these are subleading because the corresponding Regge trajectory has lower intercept.
From a string theory perspective the equations of motion for the higher spin fields dual to ${\cal O}_J$ 
should come by requiring their correspondent vertex operator to have conformal weights $(1,1)$
in the  background dual to the QCD vacuum. We shall follow an effective field theory approach, proposing a general form of the equation in a strong coupling expansion, and then
use the experimental data to fix the unknown coefficients. The proposed equation will obey two basic requirements, namely to be  compatible 
with the graviton's equation for the case $J=2$ and to reduce to the well known case in the conformal limit (pure $AdS$ space with constant dilaton).

Let us consider first the conformal case ($A(z)=\log(L/z)$ and constant dilaton). In AdS space  the spin $J$ field obeys the equation
\begin{equation}
 \left(  \nabla^2 
 - M^2  \right)  h_{a_1\cdots a_J} =0\,, \qquad \qquad
 (LM)^2=\Delta(\Delta-4) - J\,,
 \label{eq:AdS_SpinJ}
\end{equation}
where $L$ is the AdS length scale and $\Delta$ is the dimension of $\mathcal{O}_J$.
Note that this field is  symmetric, traceless 
($h^b_{\ b a_3\cdots a_J}=0$)
and transverse
($ \nabla^b h_{b a_2\cdots a_J}=0$).
This equation is invariant under the gauge transformation $\delta h_{a_1...a_J}=\nabla_{a_1}\Lambda_{a_2\dots a_J}$ with $\nabla^2\Lambda_{a_2...a_J}=0$, 
but we will modify this in such a way that this gauge symmetry will be broken, as expected for a dual of a QFT with no infinite set of conserved currents. This is trivially achieved by changing the value of $M$ in (\ref{eq:AdS_SpinJ}) 
away from the unitarity bound $\Delta=J+2$. 
The transversality condition allows us to consider  as independent components only the 
components $h_{\alpha_1...\alpha_J}$, along the 
boundary direction. These can be further decomposed into irreducible representations of the Lorentz group $SO(1,3)$, so that the traceless and divergenceless sector 
$h^{TT}_{\alpha_1 \dots \alpha_J}$ 
decouple from the rest and describe the $\mathcal{O}_J$ in the dual theory. 
Finally note that we can analyse the asymptotic form of the spin $J$ equation of motion (\ref{eq:AdS_SpinJ}) near the boundary, with the result
\begin{equation}
h_{\alpha_1\dots \alpha_J} \sim z^{4-\Delta-J} {\cal J} +...+ z^{\Delta-J} \langle {\cal O}_J\rangle +...
\label{eq:boundary_expansion}
\end{equation}
where $ {\cal J}$ denotes the source for ${\cal O}_J$. Since under the rescaling $z\rightarrow \lambda z $  the AdS field $h_{\alpha_1\dots \alpha_J}$ has dimension $J$, we conclude that the
operator ${\cal O}_J$ and its source have,  respectively, dimension $\Delta$ and $4-\Delta$, as expected.
In the case that concerns us, since QCD is nearly conformal in the UV, we can do a similar analysis near the boundary.

Next let us consider the case $J=2$, where we  have some control. This is the case of the energy-momentum tensor dual to the graviton. To describe the TT metric fluctuations we need
to assume what is the dynamics of this field.  
The simplest option is to consider 
an action for the metric and dilaton field of the form
\begin{equation}
S=M^3N_c^2\int d^5 X \sqrt{-g} \, e^{-2\Phi }\left[R +4 \left( \partial \Phi \right)^2 +V(\Phi )\right] \, , 
\label{eq:Einstein-DilatonAction}
\end{equation}
where we work in the string frame.
The  field  $\Phi$ is the dilaton without the zero mode, that is absorbed in the gravitational coupling.
This class of theories can be used to study four dimensional theories where conformal symmetry  is broken in the IR. 
To make use of the AdS/CFT dictionary one usually impose AdS asymptotics for $A(z)$, which leads to a constraint on the UV form of the potential $V(\Phi)$. This is a good approximation for large-$N_c$ QCD because it is nearly conformal in the UV
\footnote{Due to asymptotic freedom conformal symmetry is actually broken mildly in the UV by QCD logarithmic corrections.}. The way conformal symmetry is broken in the IR is determined by the potential $V(\Phi)$. As shown in \cite{gursoy_exploring_2008-1}, the confinement criteria and the spectrum of glueballs with spin $J=0,2$ constrain strongly the form of the potential $V(\Phi)$ in the IR. 

For the gravitational theory with action (\ref{eq:Einstein-DilatonAction}), the TT metric fluctuations around a background of the form (\ref{IHQCDBackground}) are given by
\begin{equation}
 \left( \nabla^2 
 -2e^{-2A(z)}  \dot{\Phi} \nabla_z 
 +2   \dot{A}^2 e^{-2A(z)}  \right) 
 h_{\alpha\beta}^{TT}=0\,.
 \label{eq:metric_fluctuations}
\end{equation}
The term with the dilaton arises because we work in the string frame;
the other term comes from the coupling of metric fluctuations to the background Riemann tensor $R_{a c b d} h^{cd}$, with
${R_{\alpha \mu \beta \nu}} = \dot{A}^2 e^{2A}\left( \eta_{\alpha\nu} \eta_{\mu\beta} -  \eta_{\alpha\beta} \eta_{\mu\nu}   \right)$
and
${R_{\alpha z \beta z}} = -\ddot{A} e^{2A} \eta_{\alpha\beta}$.
In the case of pure AdS space (\ref{eq:metric_fluctuations}) simplifies to
\begin{equation}
\left(  \nabla^2 
 - M^2  \right)  h_{\alpha\beta} ^{TT}=0\,,
\label{eq:AdSgraviton}
\end{equation}
with $(LM)^2=-2$, as expected  for the AdS graviton. We shall assume that our equation reduces to the simple form (\ref{eq:metric_fluctuations})
in the case $J=2$. Of course there could be higher order curvature corrections to this equation.
Also, in the QCD vacuum there are scalar operators with a non-zero vev that do not break Lorentz simmetry. 
The corresponding dual fields will be non-zero and may couple to the metric, just like the above curvature  and dilaton terms.  

%%%%%%%%%%%%%%%
\begin{figure}[!b]
\begin{center}
\includegraphics[height=6cm]{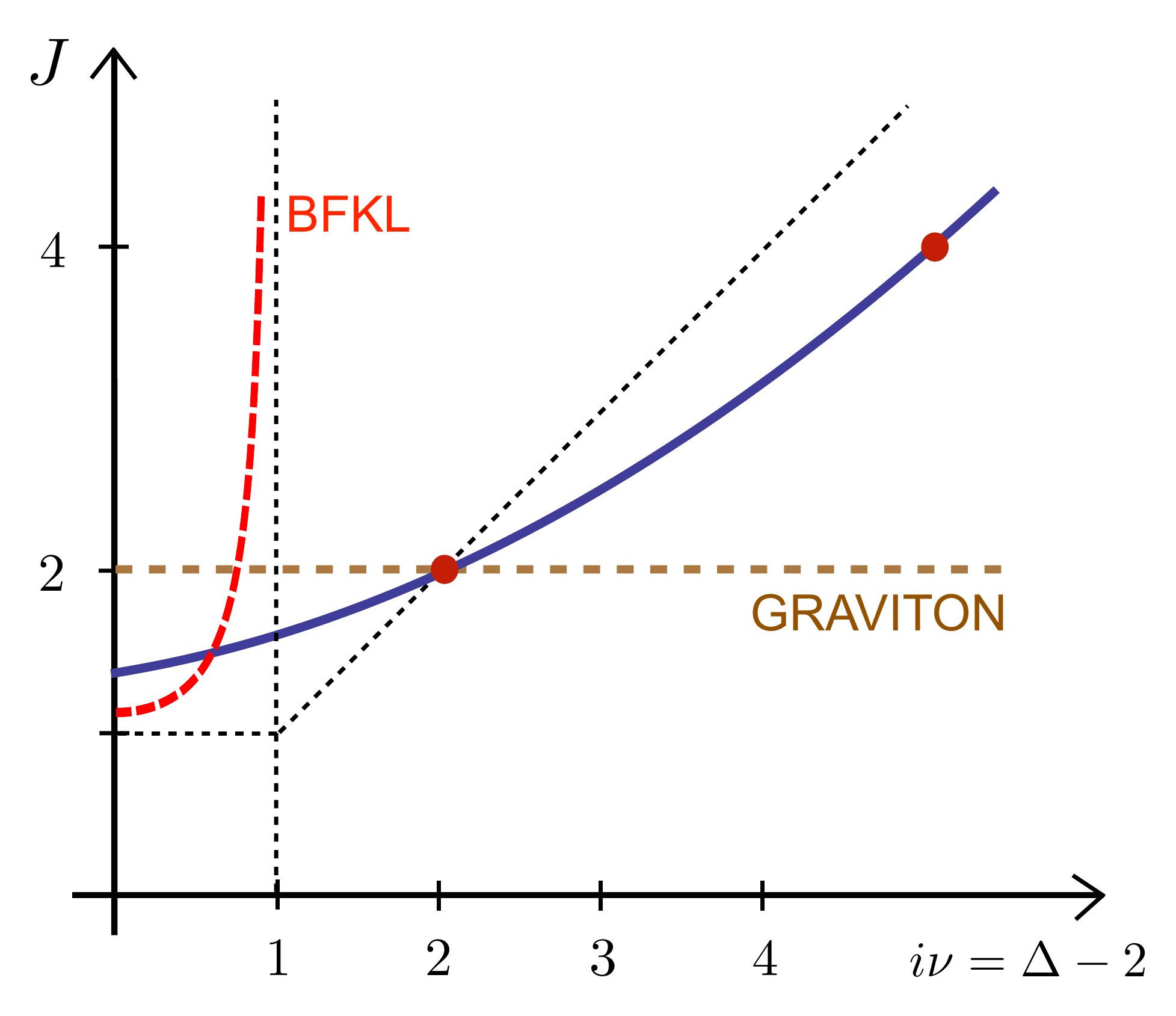}
\end{center}
\vspace{-0.5cm}
\caption{Expected form of the $\Delta=\Delta(J)$ curve (in blue). Free theory is the oblique dashed line.
At infinite coupling the curve degenerates to the graviton horizontal line (in brown). Resuming perturbation theory
one obtains, away of the spectral region, the BFKL curve  (in red).}
\label{fig:Delta(J)}
\end{figure}
%%%%%%%%%%%%%%%

Our goal is to write a two derivative equation for the spin $J$ fields using effective field theory arguments in an expansion in the derivatives of the background fields.
For this it is important to look first at the  dimension of the  operator ${\cal O}_J$, which can be written as  $\Delta=2+J + \gamma_J$, where $\gamma_J$ is the anomalous dimension. 
In free theory the operator has critical dimension $\Delta=2+J$. 
Knowledge of the curve $\Delta=\Delta(J)$ is important 
when summing over spin $J$ exchanges, since this sum is done by analytic continuation in the $J$-plane, and then by considering the region of real $J<2$.
Figure \ref{fig:Delta(J)} summarizes a few important facts about the curve $\Delta=\Delta(J)$. Let us define the variable $\nu$ by $\Delta=2+i\nu$, and consider the inverse function $J=J(\nu)$. The figure shows the perturbative BFKL result for $J(\nu)$, which is an even function of $\nu$ and has poles at $i\nu=1$. This curve is obtained by resuming $\log x$ terms in leading order perturbation theory. 
Beyond perturbation theory, the curve must pass through the energy-momentum tensor
protected point at $J=2$ and $\Delta=4$. 
We shall use a quadratic approximation to this curve that passes through this protected point,
\begin{equation}
J(\nu)\approx J_0 - {\cal D} \nu^2 =
% 2 - {\cal D}(4+\nu^2) = 
2+ {\cal D} \Delta(\Delta-4)
\,,
\qquad\qquad
4{\cal D}=2-J_0\,.
\label{eq:difusion}
\end{equation}
The use of a quadratic form for the function $J(\nu)$ is known as the diffusion limit and it is used both in
BFKL physics and in dual models that consider the AdS graviton Regge trajectory  
(see for instance \cite{costa_deeply_2012}). 

Let us now construct the proposal for the symmetric, traceless and transverse spin $J$ field $h_{a_1...a_J}$ in the dilaton-gravity 
theory (\ref{eq:Einstein-DilatonAction}). After decomposing this field in  $SO(1,3)$ irreps, the TT part
$h_{\alpha_1\cdots \alpha_J}^{TT}$ decouples from the other components and describes the 
propagating degrees of freedom. The proposed equation has the form
\begin{align}
&\bigg(  \nabla^2 
-2 \,e^{-2A} \dot{\Phi}  \nabla_z -\frac{\Delta (\Delta -4)}{L^2}  + J \dot{A}^2 e^{-2A}  +
\label{eq:dualQCD_SpinJ}
\\
&
\qquad
(J-2) e^{-2A}\left(   a \,\ddot{\Phi} +   b \left(\ddot{A}-\dot{A}^{2}\right) + c\,\dot{\Phi}^2\right) 
\bigg)  h_{\alpha_1 \dots \alpha_J}^{TT}=0\,,
\nonumber
\end{align}
where $a,b$ and $c$ are constants.
Several comments are in order:
(i) For $J=2$ this equation reduce sto the graviton equation (\ref{eq:metric_fluctuations});
(ii) In the AdS case all terms in the second line vanish and the equation reduces to (\ref{eq:AdS_SpinJ}) for the TT components;
(iii) The second term comes from the tree level coupling  of a closed string, as appropriate for the  graviton Regge trajectory in a large $N$ approximation;
(iv) This action contains all possible terms of  dimension inverse squared length compatible with constraints (i) and (ii) above. Notice that the 
term $\dot{\Phi}\dot{A}$ is absent because it reduces to other two derivative terms of $A$ and $\Phi$ by the equations of motion.
Also, note that the terms with two $z$ derivatives are  accompanied by a metric factor $g^{zz}=e^{-2A}$ from covariance 
of the 5-d theory. The exception is the first term, which itself includes the 5-d metric $\nabla^2=g^{ab}\nabla_a\nabla_b$, and the third
that is a mass term related to the dimension of the dual operator, which requires a length scale $L$. 

It is important to realize that (\ref{eq:dualQCD_SpinJ}) is not supposed to work for any $J$. Instead, we are building the analytic continuation 
of such an equation, which we want to use around $J=2$. We expect this to be the case for large coupling,  which is the case for the dense gluon medium observed in 
the low $x$ regime.
In practice, we will look at the first pomeron poles that appear between $0.6\lesssim J\lesssim1.5$ (for $t=0$, as required in the computation of the 
total cross section). This justifies why we left the coefficients $a,b,c$ in the second line of (\ref{eq:dualQCD_SpinJ}) constant and consider only the first term in the
$J$ expansion around 2.

Finally let us consider the third term in (\ref{eq:dualQCD_SpinJ}). This mass term is determined  by the analytic continuation of the dimension of the exchanged operators $\Delta=\Delta(J)$.
We will write the following formula
\begin{equation}
\label{eq:diff_limit}
\frac{\Delta(\Delta - 4)}{L^2}= \frac{2}{l_s^2}\,(J-2) \left( 1 + \frac{d}{\sqrt{\lambda}} \right) + \frac{1}{\lambda^{4/3}} (J^2 -4)\,,
\end{equation}
where $\lambda=e^\Phi$ is the 't Hooft coupling, $d$ is a constant and $l_s$  is a length scale set by the QCD string, which will be one of our phenomenological parameters. The first term follows directly from the diffusion limit  (\ref{eq:difusion}), relating the
scales $L$  and $l_s$ via  ${\cal D}$. The diffusion limit is a strong coupling expansion, so it is natural that the dimension of the operators gets corrected in an expansion in $1/\sqrt{\lambda}$.
This is the reason for adding the second term in (\ref{eq:diff_limit}), following exactly what happens in ${\cal N}=4$ SYM \cite{costa_conformal_2012,Cornalba:2007fs}. 
This term can be added to correct the IR physics, but it is still subleading in the
UV, when compared with the last term. The effect of this correction is to make the scale $l_s$ dependent of the energy scale, while keeping the general shape of 
curve $\Delta=\Delta(J)$ in figure \ref{fig:Delta(J)}. 
The last term in (\ref{eq:diff_limit}) was added simply to reproduce the correct free theory result that is necessary to be obeyed near the boundary in the UV.
More concretely, in order to obtain a scaling of the form  (\ref{eq:boundary_expansion}), with  the free dimension $\Delta=J+2$, we need this last term. This follows by considering the asymptotic value
of the background fields and then analysing our spin $J$ equation near the boundary to obtain $h_{+ \dots +}^{TT} \sim z^2$. 
This behaviour is important since it implies Bjorken scaling at the UV. We can regard (\ref{eq:diff_limit}) as an interpolating function between the IR and UV that matches the expected
form of the dimension of the spin $J$ operator in both regions. This is the same type of approach followed in phenomenological holographic QCD models.

To sum up, we shall consider the effective Reggeon equation 
(\ref{eq:dualQCD_SpinJ}), with (\ref{eq:diff_limit}),  to describe the exchange  of all the spin $J$ fields in the graviton Regge trajectory. This equation contains 5 parameters that will be fixed by the data, namely
the constants $a,b,c,d$ and $l_s$. 

We finish the analysis of the spin $J$ equation with a remark. In the same lines of \cite{karch_linear_2006} we 
can try to write a quadratic effective action for the spin $J$ symmetric, traceless and transverse field, such that its $SO(1,3)$ irrep TT part 
obeys the proposed free equation.  Such an action would have the form
\begin{equation}
I = \frac12 \int d^5 X \sqrt{-g}\, e^{-2 \Phi} \Big [ \nabla_b h_{a_1 \dots a_J} \nabla^b h^{a_1 \dots a_J} - M^2(z) h_{a_1 \dots a_J} h^{a_1 \dots a_J} + \dots \Big ] \, ,
\label{HigherSpinAction}
\end{equation}
where the dots represent terms quadratic in $h_{a_1 \dots a_J}$ that are higher in the derivatives of either $h_{a_1 \dots a_J}$ or the background fields.
Since in the QCD vacuum only scalars under the $SO(1,3)$ irrep decomposition are allowed to adquire a vev, the mass term in  (\ref{HigherSpinAction})
includes all such possibilities. We are also treating the dilaton field in a special way, by allowing a very specific coupling in the overall action. In particular, other scalar 
fields could also have a non-trivial coupling to the kinetic term
\footnote{Since we write a 5-d action, one could also have fields with a vev proportional to the 5-d metric $\eta_{ab}$. An example is the 
background Riemann tensor that  can couple to the spin $J$ field (for instance, the metric fluctuations do). However, for 
traceless fields only mass terms of the type  written in (\ref{HigherSpinAction}) will survive.}.
It is simple to see that our proposal (\ref{eq:dualQCD_SpinJ}), with (\ref{eq:diff_limit}), corresponds to setting 
\begin{equation}
M^2(z) =- J  \, e^{-2A} \dot A^2 +m^2(z) \,,
\end{equation}
with
\begin{equation}
    \label{eq:final_potential}
    m^2(z)= (J-2) \left[ 
        \frac{2}{l_s^2} \left( 1 + \frac{d}{\sqrt{\lambda}} \right) + \frac{J+2}{\lambda^{4/3}} + 
     e^{-2A}\left(   a \,\ddot{\Phi} +   b \left(\ddot{A}-\dot{A}^{2}\right) + c\,\dot{\Phi}^2\right) \right].
\end{equation}

%%%%%%%%%%%%%%%%%%%%%%%%%%%%%%%%%%
\subsection{Effective Schr\"{o}dinger problem}
%%%%%%%%%%%%%%%%%%%%%%%%%%%%%%%%%%

The amplitude ($\ref{eq:witten_diagram}$) computes the leading term of the 
Witten diagram describing the exchange of the spin $J$ field in the Regge limit, whose propagator obeys the equation
\begin{equation}
( {\cal D} \Pi)_{a_1\cdots a_J,b_1\cdots b_J}(X,X') =
 i e^{2\Phi}g_{a_1\left(b_1\right.} \!\cdots g_{|a_J| \left. b_J\right)} \delta_5(X,X') - {\rm traces}\,,
\label{eq:SpinJpropagator_eq}
\end{equation}
for some  second order differential operator ${\cal D}$ whose action on the $TT$ part of the spin $J$ field is defined by  (\ref{eq:dualQCD_SpinJ}).
For Regge kinematics, however, we are only interested in the component  $\Pi_{+\dots+,-\dots-}$ of the propagator, in
the limit where the exchanged momentum has $q_+ =O(1/\sqrt{s})\sim0$, as can be seen from the  kinematics of the external photons (\ref{eq:momenta_photon}).
Thus, we can take $\partial_{+} h^{TT}_{+ \dots +} = 0$, which implies that the $+\dots+$ component of (\ref{eq:dualQCD_SpinJ}) decouples from the 
other components, taking the following form \footnote{For example, the bulk Laplacian projected in the boundary indices gives $\nabla^2 h_{\alpha_{1}...\alpha_{J}} = \left(e^{JA\left(z\right)}\nabla_{0}^{2}e^{-JA\left(z\right)}-JA'\left(z\right)^{2}e^{-2A\left(z\right)}\right) h_{\alpha_{1}...\alpha_{J}}+O(1/\sqrt{s})$, where $\nabla_{0}^{2}$ is the bulk scalar Laplacian.}
\begin{equation}
\Big \{  \left [ \partial_z + 2 \dot A - 2 \dot \Phi \right ]  \left [ \partial_z - \dot A \right ] + \nabla_\perp^2 - m^2(z) e^{2A} \Big \} e^{(1-J)A} h^{TT}_{+ \dots +} = 0 \, . 
\label{TTEq3d}
\end{equation}
This equation can be re-casted as a 1-d quantum mechanics problem, that is, setting 
\begin{equation}
h^{TT}_{+ \dots +} =  e^{i q \cdot x} e^{(J-1)A} e^{B(z)} \psi(z) \, , 
\end{equation}
with $q_{+} = 0$,
and choosing  $B(z) =  \Phi-A/2$ to cancel the term linear in the derivative $\partial_z$, 
equation (\ref{TTEq3d}) takes the Schr\"{o}dinger form
\begin{eqnarray}
\Big[ \partial_z^2 +t - V(z) \Big] \psi (z) = 0  \, , \label{TTEqSchrodinger}
\end{eqnarray}
where $t = - q_\perp^2$ and the potential $V$ is given by
\begin{equation}
    V(z) = \frac32 \left (\ddot A - \frac23 \ddot \Phi \right ) + \frac94 \left (\dot A - \frac23 \dot \Phi \right )^2 + m^2(z) e^{2A}\,.
\end{equation}
The energy spectrum for each integer $J$  quantises $t=t_n(J)$, therefore yielding the
glueball  masses (although we only expect the proposed equation to be a good approximation for analytically continued values of $J$ around $J=2$, and certainly not in the 
asymptotic regime of large $J$). 
As expected, for $J=2$ this potential reduces to the one obtained from linearized Einstein equations, since $m^2(z) = 0$ for $J=2$.

Finally we can consider the integrated propagator $G_J(z,z',l_\perp)$ defined in (\ref{eq:transverse_prop}). This is the scalar propagator
obtained from integrating the  component  $\Pi_{+\dots+,-\dots-}$ of the full propagator along the light-rays. From the differential equation
(\ref{TTEq3d}) if follows that
 \begin{equation}
\left[ \Delta_3 - e^{-2A(z)}  \left( 2 \dot{\Phi} \partial_z + 2\dot{A}^2 + \ddot{A} -2\dot{A}\dot{\Phi}  \right) 
  - m^2(z)\right]G_J(z,z',l_\perp)=- e^{2\Phi} \delta_3(x,x')\,,
  \label{eq:Gprop_eq}
\end{equation}
where here  $x=(z,x_\perp)$ and $\bar{x}=(\bar{z},\bar{x}_\perp)$ are points in the scattering transverse space with metric 
$ds_{3}^2 = e^{2A(z)} \big[ dz^2 + dx_\perp^2 \big]$, and $\Delta_3$ is the corresponding Laplacian.
It is now clear that writing
\begin{equation}
G_J(z,\bar{z},t)= e^{B(z)} \psi(z)\,,
\end{equation}
the homogeneous solution to (\ref{eq:Gprop_eq}) is exactly given by the Schr\"{o}dinger problem 
of (\ref{TTEqSchrodinger}). 
Moreover, using the spectral representation $\sum_n \psi_n(z)\psi^*_n(\bar{z})=\delta(z-\bar{z})$, 
we conclude that
\begin{equation}
G_J(z,\bar{z},t)= e^{B(z)+B(\bar{z})} \sum_n  \frac{\psi_n(z)\psi^*_n(\bar{z})}{t_n(J)-t}\,.
\end{equation}
This result was used in subsection \ref{subsec:Reggepoles} to derive the contribution of Regge poles to the DIS structure functions. Notice that both
the eigenvalues $t_n$ and the functions $\psi_n$ depend on $J$.

%%%%%%%%%%%%%%%%%%%%%%%%%%%%%%%%%%%%%%%%%%%
\section{Fit of DIS data in IHQCD}
\label{sec:fit}
%%%%%%%%%%%%%%%%%%%%%%%%%%%%%%%%%%%%%%%%%%%

In this section we will test the previous phenomenological model for DIS against the combined H1-ZEUS data points for $x<0.01$ from \cite{Aaron:2009aa}, as shown in figure \ref{fig:HERAdata}.  
We will look for the optimal values of the free parameters in the structure function $F_2(x,Q^2)$ given in (\ref{F1F2Regge}). This function depends on the couplings $g_n$ 
to each Reggeon, given by (\ref{eq:couplings}), and on the parameters $l_s$, $a$, $b$, $c$ and $d$ in (\ref{eq:final_potential}) that characterize the analytic continuation of the graviton Regge trajectory. 
At the end of the day we will fix the shape of the first glueball    Regge trajectories, which then can be compared with  the known higher spin glueball data obtained from lattice computations. 
Not only we are able to fit DIS data, we shall see that our results are also compatible with the most recent results we have found so far for the higher spin glueball spectrum \cite{meyer_glueball_2005,meyer_glueball_2005-1}. 
This is expected, since the spectral and scattering data ought to be  connected consistently.

Although we kept in the previous sections our discussion of  pomeron physics general, we need to use a  specific holographic QCD model to test our ideas. As in our previous work \cite{Ballon-Bayona:2015wra}, we shall consider 
the Improved Holographic QCD (IHQCD) model proposed in \cite{gursoy_exploring_2008-1}. The QCD vacuum is described by a  dilaton gravity theory with an action of the form (\ref{eq:Einstein-DilatonAction}). The potential $V(\Phi)$ is then judicially 
chosen such that the theory reproduces the QCD beta function in the UV and confines in the IR. In our fit we will consider data points with $Q^2$ as large as $400\  {\rm MeV}$. For this reason
we need to start close to the AdS boundary at   $z_{min}\equiv e^{-A_0}=0.0067$ with $\lambda_0= 0.0337462$. These initial values of $A(z)$ and $\lambda(z)$ are consistent with the choice of $\Lambda_{QCD}=0.292GeV$ which gives the value of the lowest mass of the spin 2 glueball proposed originally in \cite{gursoy_exploring_2008-1}.
For the maximum value of the holographic variable we chose $z_{max}=6$. Eventually we have changed $z_{max}$ to a bigger value, the results showing no sensitivity.
\footnote{For the interested reader we have release our spectral code under a MIT License, which you can find at github \url{https://github.com/rcarcasses/schrodinger}.}

\begin{figure}[t!]
    \centering
    \includegraphics[scale=0.7]{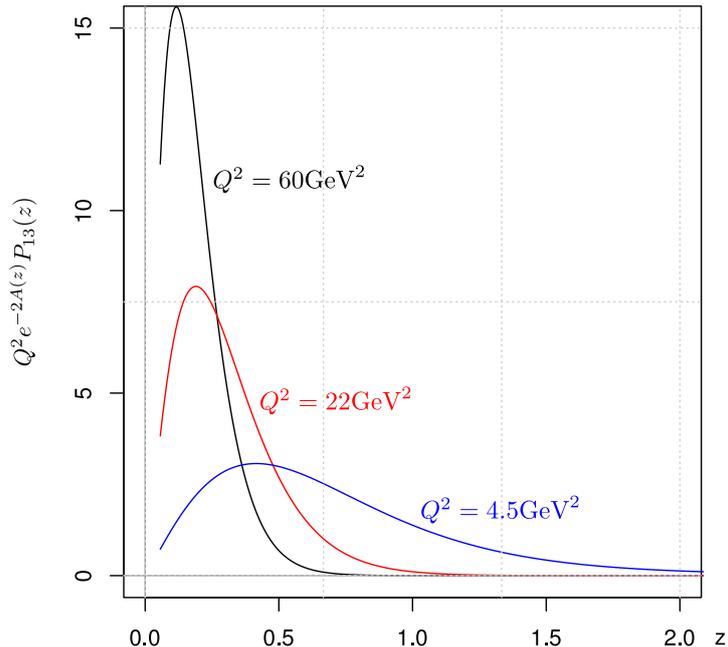}
    \caption{The function previously approximated to a delta function in section \ref{sec:pheno}. The quality of the approximation clearly decreases as $Q^2$ become smaller.}
    \label{fig:delta_func_app}
\end{figure}

Next we need to compute the non-normalizable modes associated to the current operator sourced by the off-shell photon, as explained in section \ref{ReggeTheory}  (equations (\ref{eq:ansatzA}) and (\ref{eq:eqsA})).
This is done by solving the equation for the $U(1)$ gauge field in the bulk for each of the $Q^2$ available in the data.
The dependence of the structure function $F_2(x,Q^2)$ on the external probe arises from the shape of the function $P_{13}{}^{(2)}\!\left(Q^2,z\right)$ defined in (\ref{eq:P13}), since this function is 
then integrated along $z$ in (\ref{eq:P13bar}).  In figure  \ref{fig:delta_func_app} we plot the function $P_{13}{}^{(2)}\!\left(Q^2,z\right) Q^2 e^{-2A}$
 for several values of $Q^2$. 

At this point we can confirm that the approximation of the external photon wave functions
to the integral (\ref{eq:P13bar}) by a Dirac delta function, as assumed in section \ref{sec:pheno}, only works for large values of $Q^2$.
Writing  the integrand in (\ref{eq:P13bar}) as 
$P_{13}{}^{(2)}\!\left(Q^2,z\right) Q^2 e^{-2A}  \times function(z)$, it is simple to see that the first function behaves as a delta function 
for large $Q$, while the $function(z)$ is smooth enough such that the integral gets no contribution from the boundary at $z=0$.
This is the reason why we kept our discussion of section \ref{sec:pheno} at a more qualitative level, since in the reconstruction of
the wave functions in figure  \ref{fig:exp_wf} we did the replacement $z \sim 1/Q$.
Clearly such replacement is a gross approximation for many of the values of $Q$ we are considering. \footnote{
In fact it is puzzling how such approximation, which  leads to a closed formula for $F_2$ using a hard wall model  \cite{brower_string-gauge_2010}, works very well with results 
comparable to those presented in this paper.}
An honest computation would  involve solving numerically an integral equation for the $\psi_{0,1}$ given the experimental $f_{0,1}(Q)$, with a kernel of the form shown in figure \ref{fig:delta_func_app}. 
Nevertheless we would expect that this correction will not introduce any new extreme point for the wavefunction but to deform it in a non trivial way in the region where the delta function approximation is bad, giving still the right number of nodes for each $\psi$, which is really the point we wanted to emphasize in section \ref{sec:pheno}.

Let us remark that, in contrast with the external off-shell photon with varying virtuality $Q^2$,  the target proton wave function does not require a detailed description  of the holographic dual $\Upsilon(P^2,z)$. The dependence on that normalizable mode was carried by $P_{24}(P^2,z)$ and absorbed in the coupling constants $g_n$, as shown in (\ref{eq:couplings}).

Following our program we define an error function depending of the phenomenological parameters ${\alpha_i}$. 
This defines a optimization problem where we wish to find the values of ${\alpha_i}$  that minimize the quantity
\begin{equation}
    \chi^2\equiv\sum_{k=1}^N\left( \frac{F_2(Q_k^2,x_k;{\alpha_i})-F_2^{exp}(Q_k^2,x_k)}{\sigma_k} \right)^2,
\end{equation}
which is just a weighted least square fit where the weight is the inverse of the error in the measurement, such that quantities with bigger error affect less the result. A widely accepted criteria for the quality of a fit is that the quantity $\chi^2_{d.o.f.}\equiv\chi^2/(N-N_{par})$, where $N_{par}$ is the number of parameters to be fitted, is closer to one.

%%%%%%%%%%%%%%%%%%%%%%%%%%%%%%%%%%%%%%%%%%%
\subsection{The fit}
\label{sec:results}
%%%%%%%%%%%%%%%%%%%%%%%%%%%%%%%%%%%%%%%%%%%

We proceed to find the optimal values for the phenomenological parameters. 
As explained in the introduction and shown in figure \ref{fig:HERAdata} we consider data points with $x<0.01$ to be in the limit of  Regge kinematics.
Let us remind ourselves that in our derivation we have dropped terms of order $1/s$. Such terms are sub-leading with respect to the first trajectories as long as their intercept does not differ from the leading one at most by unit. This validates our choice of retaining the first daughter trajectories. 

We have found that with only two trajectories it is possible to provide a good fit for the DIS data, but unfortunately the second intercept does not correspond to the soft pomeron. As explained before, it is desirable that the second intercept matches that of the soft pomeron given the experimental evidence from total cross sections of  soft probes. 
We have also found that fixing the intercept of the second trajectory to 
$1.08\sim 1.09$, and performing the fit with only two Pomerons, does not provide a good fit. 

\begin{figure}[!t]
    \centering
    \includegraphics[scale=0.9]{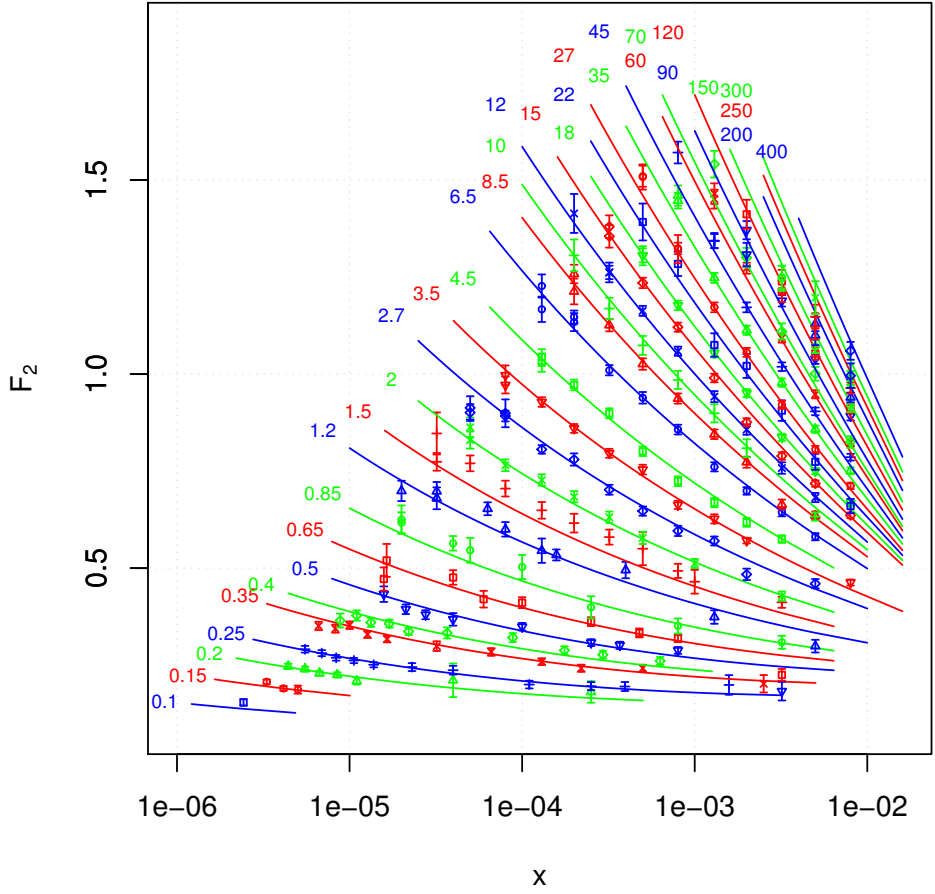}
    \caption{Plot of the experimental $F_2$ data versus   the prediction of our model. We cover a very large kinematical window with $x<0.01$ and $0.1<Q^2<400$ in ${\rm GeV}^2$, in a total of $249$ points.  
    The $\chi^2_{d.o.f.}$ of this fit is $1.7$.}
    \label{fig:F2fit}
\end{figure}

Thus the most reasonable thing to do is to fix the soft pomeron intercept and include a third or a fourth trajectory in our fit, similar to \cite{donnachie_new_2001}, but in our case these trajectories are associated to 
glueballs instead of mesons. The reason we stopped at the fourth trajectory is that it is enough to obtain a good quality fit. We found that these trajectories still have an intercept above the mesons one , i.e. $j_n > 0.55$,
so they are important and should be taken into  account\footnote{This is subtle: if the meson trajectories bend in the same way as the glueball ones then their intercepts will raise. We plan to analyse mesons' contribution in a future work.}. Figure \ref{fig:F2fit} shows our best fit from which we obtained  $\chi^2_{d.o.f.}=1.7$.  The corresponding parameters 
are listed in the table 1 and more details can be found in the appendix \ref{sec:on_numerics}.
 
\begin{table}[!t]
    \label{tab:parameters}
    \centering
    \begin{tabular}{|c||c||c|}
        \hline 
         Pomeron equation coefficients & coupling  & Intercept\\
        \hline 
        \hline 
			 $\ \ a = -4.35$ & $g_{0}=0.175$ &$j_0=1.17$\\
        \hline 
			 $b= 1.41$ & $g_{1}= 0.121$  &$j_1=1.09$\\
        \hline 
			 $\ \,c= 0.626$ & $g_{2}= 0.297$   &$\ j_2=0.969$ \\
        \hline 
			 $\ \ \ \,d= -0.117$ & $\ g_{3}= -1.63$  &$\ j_3=0.900$\\
         \hline 
			 $\, l_{s} = 0.153$   &--  & --\\
        \hline 
    \end{tabular}
        \caption{The nine parameters for our best fit. As an output we also show the intercept of the  first four pomeron trajectories (in fact we forced the second trajectory to have the soft pomeron value, so only the
        other values  are a prediction of the model).}
\end{table}

\begin{figure}[t!]
    \centering
    \includegraphics[scale=0.8]{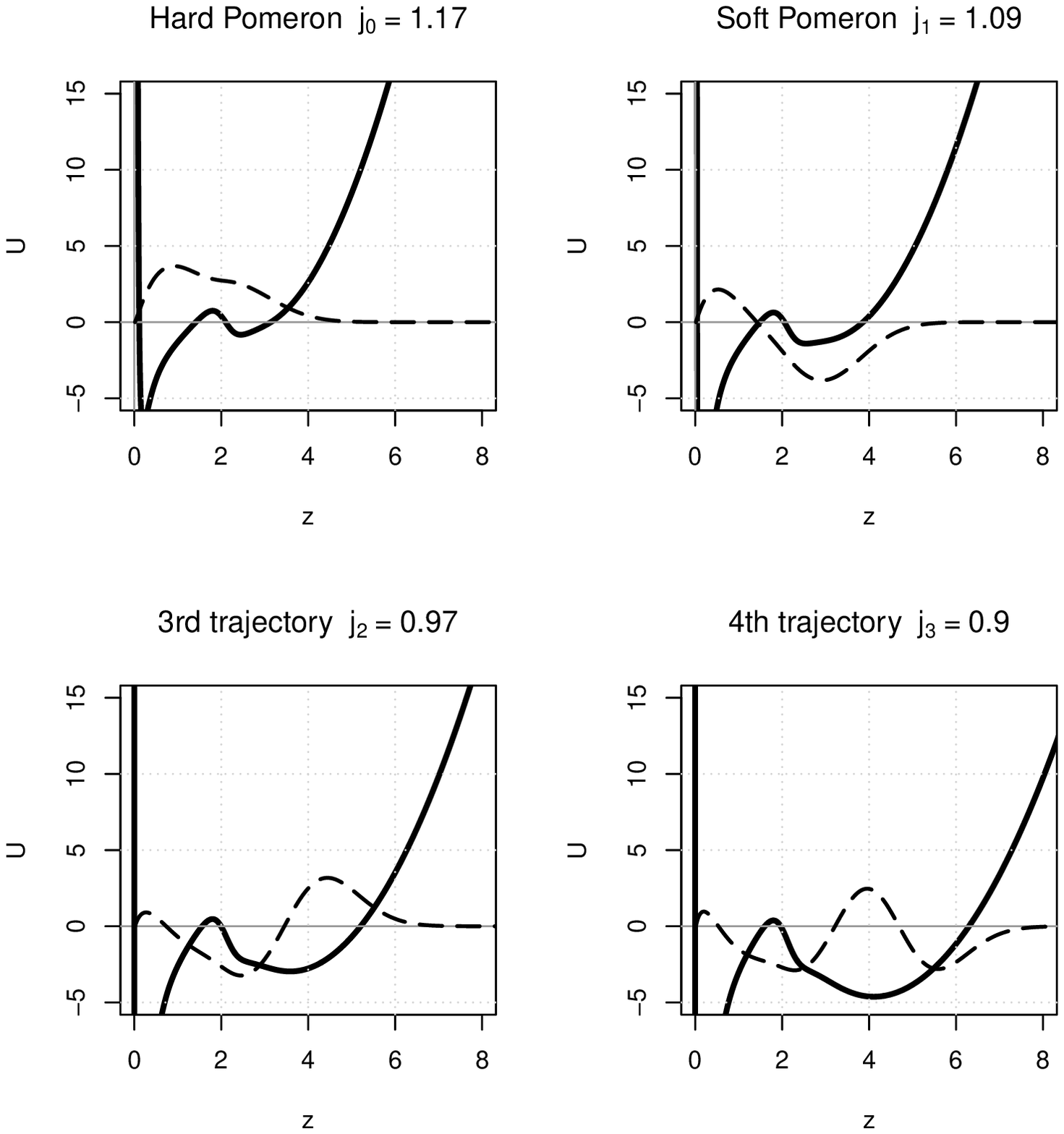}
    \caption{Wavefunctions and corresponding potential  for optimal phenomenological values for the hard and soft  pomerons, and  also for the other daughter trajectories considered in the fit. 
    The normalized wavefunctions have been scaled by 5 in this figure.}
    \label{fig:spectrum}
\end{figure}

In figure  \ref{fig:spectrum} we show, for each pomeron, the wave function and corresponding  potential in the associated Schr\"{o}dinger problem. As anticipated the wavefunctions resemble those of section \ref{sec:pheno} with some sort of deformation, specially for large $z$, due to the use of the right functions for the external off-shell photon.
Recall that, as we vary the spin $J$ in the Reggeon equation (\ref{TTEqSchrodinger}), the intercept of the $n$-th trajectory is given by the value of $J$ for which the energy of the  $n$-th excited state crosses zero. 
Thus the wave function shown in each figure is the zero energy state for the potential shown in the same figure. We see that as $J$ decreases the potential spreads to
 the IR region. This fact is at the heart of the  decrease of the effective intercept with a decreasing virtuality $Q^2$ shown in figure \ref{fig:intercept} in the introduction. In other words, as the process becomes more localized in the IR the 
 wave function of the hard-pomeron is very suppressed, leading to more important contributions from the other daughter trajectories. The potential also shows a very sharp minimum near the UV, that localizes the
 hard pomeron wave function near the boundary. However, we do not fully understand the two minima behaviour exhibited by the potential. For instance, it could be that this is just an artifact of the 
 specific interpolation between the IR and UV physics in the holographic QCD background considered here. Nevertheless, we believe the most important fact here 
 is that the wave functions are smooth and are more spread across  the IR in the  case of the daughter trajectories. 

%%%%%%%%%%%%%%%%%%%%%%%%%%%%%%%%%%%%%%%%%%%
\subsection{Regge trajectories}
\label{sec:results}
%%%%%%%%%%%%%%%%%%%%%%%%%%%%%%%%%%%%%%%%%%%

The power of Regge theory relies on the fact that spectrum and scattering physics are related in a very natural way. 
In the conventional approach one organizes the spectrum in Regge trajectories $J=j(M^2)$, for integer  $J$. 
Then one considers processes where particles in a given trajectory $j_n(t)$ are exchanged. Regge 
theory  predicts that the contribution of the trajectory $j_n(t)$ to the cross section behaves as $s^{j_n(t)}$, where the function $j_n(t)$  is analytically continued to negative $t$.  

Here we are following a similar strategy. First we construct a phenomenological model and then fix the unknown coefficients by confronting the model to scattering data. In fact, since we consider a total cross
section for an inelastic process, the scattering data we used  is directly related to the value of each Regge trajectory at $t=0$. We can then look how each Regge trajectory behaves for positive and negative $t$. 
The plot of the first four pomeron Regge trajectories considered in this work was presented in figure \ref{fig:trajectories} in the introduction.
It is rewarding to see that the  first two trajectories, that is the hard and soft pomeron, pass reasonably well through the   
known lattice QCD data for the masses of the higher spin glueballs. One should keep in mind, though, that we are using a phenomenological holographic QCD model, and also that this lattice data refers to pure glue with SU(3)
gauge group.

An interesting feature of our leading Regge trajectories is that they coincide in shape with what was recently proposed to be the universal behaviour for weakly coupled theories with massive higher spin fields 
\cite{caron-huot_strings_2016}, where the authors argue that $j(t)\sim t+...$ for large positive $t$ and that $j(t)=const$ for large negative $t$. Another interesting fact of figure \ref{fig:trajectories} is that the trajectories are very close to each other in the scattering region of $t$<0. We comment on possible implications to  elastic differential cross sections of soft probes in the conclusions.

%%%%%%%%%%%%%%%%%%%%%%%%%%%%%%%%
\section{Conclusions}
\label{sec:con}
%%%%%%%%%%%%%%%%%%%%%%%%%%%%%%%%

In this paper we have shown how holographic QCD can be effectively used to address an essentially non-perturbative problem in QCD, that of the pomeron Regge trajectories. The construction is general, but to test it against experimental data we have considered the Improved Holographic QCD background proposed in \cite{gursoy_exploring_2008,gursoy_exploring_2008-1,gursoy_improved_2011}. 
More specifically we have been able to explain satisfactory DIS data in the $x<0.01$ region, covering a large region for the photon virtuality $Q^2$. Moreover, the same Regge trajectories that describe DIS data are compatible with the lattice data for the higher spin glueball spectrum. 

There is a natural parameter one could have chosen to tune, which is $\Lambda_{QCD}$. In IHQCD this is equivalent to choose some $A_0$ and $\lambda_0$ at a given value of $z=z_0$. This parameter was left fixed to the same value the authors of \cite{gursoy_exploring_2008,gursoy_exploring_2008-1} suggest, since in the original papers it was fixed such that the mass of the scalar glueball coincides with that of  lattice QCD. In our case changing it would lead simply to a rescaling of all dimensionful quantities in the model, like for instance the unknown coefficients we were fitting, and it has the effect of shrinking/expanding the $t$ axis of figure \ref{fig:trajectories}. Since we are also confronting our model with spectral data, we decided not to change that number. At most we could match exactly the mass of the lightest  spin two glueball, but our hard pomeron trajectory already passes very close to that point as can be see from figure \ref{fig:trajectories}, so we decided not to use such extra freedom.

Our works points towards the solution of a long standing problem in QCD, namely the nature of the hard and soft pomeron. In our framework both arise as distinct Regge trajectories made of glueballs. In the dual picture they 
originate from the graviton trajectory, which degenerates in many trajectories once it is quantized in the asymptotically AdS space (which can be thought as a gravitational box).
The way these trajectories appear in DIS data, by means of a wave function of a Schr\"{o}dinger problem, clearly calls for a reconstruction of the holographic dual of QCD.  
Somehow this is what we have done for the graviton Regge trajectory associated with higher spin glueballs. We considered a holographic QCD model that describes the QCD vacuum, and then used effective field theory arguments to reconstruct the analytic continuation of the spin $J$ equation of motion that best fits the data.

%One possible undesired aspect of our work is the bump observed %in the pomeron effective potential, as can be seen in figure \ref{fig:spectrum}. We do not have much to say about these bumps, except that
%they could be an artefact of our phenomenological model which interpolates between the strongly coupled physics in IR and  the weakly coupled one in the UV. What is important here is that the potential 
%becomes more spread to the IR as we decrease $J$. Such behaviour is determined by the  spin $J$ equation in a region where it is more justifiable, so we expect this important property to survive in 
%a more realistic holographic model.

An important point that pops up from the analysis of figure \ref{fig:trajectories} is that eventually meson trajectories will also contribute to the scattering (either in DIS, as we vary the virtuality $Q^2$, 
or in differential cross sections for elastic processes, as we vary $t$).  
This was well noticed in the work of Donnachie and Landshoff and we expect that including the dynamics of the higher spin fields dual to the mesons might improve the quality of our fit.

Requiring several trajectories to explain the DIS data is also compatible with the picture of having a branch cut structure in the $J$-plane that turns into a set of poles due to the breaking of conformal invariance.
In fact, DIS data was also successfully reproduced using a hardwall model with a conformal pomeron \cite{brower_string-gauge_2010}. Moreover, the perturbative approach that uses the BFKL pomeron also breaks conformal 
invariance and then considers several daughter trajectories \cite{kowalski_using_2010}. However, in that case one needs to consider a very large number of trajectories, leading to a very large number of free parameters in the model. We believe holographic QCD is better suited to address pomeron physics, because the whole construction is better suited to study strongly coupled phenomena. 

An interesting direction to pursue is to attempt to explain elastic differential cross-sections of soft probes, which are determined by the Regge trajectories in the small negative $t$ region. 
We expect that the leading trajectory will be suppressed, because the hard-pomeron wave function is more localized around the UV region. However, as we move in $t$ the trajectories are very close to each other which brings the possibility of observing interference between them. For instance, for $pp$ scattering  there is a dip observed around $t\sim -1 {\rm GeV}^2$. Current approaches which have attempted to explain this data using linear trajectories suggest that terms beyond single Reggeon exchange are needed, such as 
eikonal or triple gluon exchange inspired terms, to perform a good fit, like for example in  \cite{donnachie_elastic_2011}. We believe that since our trajectories are non-linear and slightly fluctuating in that kinematical region, they could also lead to an explanation of those interesting features without further considerations.

%%%%%%%%%%%%%%%%%%%%%%%%%%%%%%%
\acknowledgments
%%%%%%%%%%%%%%%%%%%%%%%%%%%%%%%
The authors are grateful to Marko Djurić for his valuable work during the early stages of this project. We also benefited from discussions with Jo\~ao Penedones.
This research received funding from the [European Union] 7th Framework Programme (Marie Curie Actions) under grant agreement 317089 (GATIS), from the grant CERN/FIS-NUC/0045/2015 and from the 
Simons Foundation grant 488637 (Simons collaboration on the Non-perturbative bootstrap). 
The work of A.B-B is funded by S\~ao Paulo Research Foundation (FAPESP) under the grant 2015/17609-3. Centro de F\'\i sica do Porto is partially funded by the Foundation for Science and Technology of Portugal (FCT).

%%%%%%%%%%%%%%%%%%%%%%%%%%%%%%%
\appendix
\section{$U(1)$ field in holographic QCD}
\label{U(1)field}
%%%%%%%%%%%%%%%%%%%%%%%%%%%%%%%

As explained in subsection \ref{ReggeTheory}, the five dimensional gauge field dual to a $U(1)$ current is described by the action (\ref{Maxwell}). The corresponding field equations are 
\begin{equation}
\partial_b\left[\sqrt{-g} \, e^{-\Phi }F^{ba}\right]=0 \, .\label{Maxwelleqs}     
\end{equation}
It is convenient to split these equations in $(z,x^\mu)$ components :
\begin{align}
&\Box A_z - \partial_z( \partial_{\hat \mu} A^{\hat \mu} ) = 0\, , 
\label{Maxwelleqsv2} \\
&
e^{\Phi-A} \partial_z \left [ e^{A - \Phi} \partial_z A^{\hat \mu} \right ] + \Box A^{\hat \mu} - \partial^{\hat \mu} \left [ e^{\Phi-A} \partial_z (e^{A - \Phi} A_z ) + \partial_{\hat \nu} A^{\hat \nu} \right ] = 0  \,, 
\nonumber
\end{align}
where $\hat \mu$ is raised with $\eta_{\mu \nu}$ and $\Box = \eta^{\mu \nu} \partial_\mu \partial_\nu$. 

Decomposing the gauge field $A^{\hat \mu}$ into its divergenceless and divergenceful parts, i.e. 
\begin{eqnarray}
A^{\hat \mu} = A^{\hat \mu}_{\perp} + \partial^{\hat \mu} \phi \, ,  
\qquad\qquad (\partial_{\hat \mu} A^{\hat \mu}_{\perp} = 0) \, ,
\end{eqnarray}
 the field equations (\ref{Maxwelleqsv2}) reduce to 
\begin{align}
&e^{\Phi-A} \partial_z \left [ e^{A - \Phi} \partial_z A^{\hat \mu}_{\perp} \right ] + \Box A^{\hat \mu}_{\perp} = 0 \, , 
\nonumber\\
&A_z = \partial_z \phi \,. 
\end{align}
Now consider the Lorentz-like gauge 
\begin{equation}
e^{\Phi-A} \partial_z (e^{A - \Phi} A_z ) + \partial_{\hat \nu} A^{\hat \nu} = 0 \, . \label{gaugechoice} 
\end{equation}
Under that gauge $\phi$ satisfies same equation as $A^{\hat \mu}_{\perp}$.
For a plane-wave ansatz we get
\begin{align}
A_{\mu }(x,z)&=\xi _{\mu} \, e^{\text{iq}\cdot x}f\left(Q^2,z\right) \, , 
\nonumber\\
A_z(x,z)&=e^{\text{iq}\cdot x}g\left(Q^2,z\right) \, ,
\end{align}
where $f$ and $g$ satisfy the equations
\begin{align}
&e^{\Phi -A}\partial _z\left(e^{A-\Phi }\partial _zf\right) - Q^2f=0 \, , 
\nonumber\\ 
&g=-i\frac{q\cdot \xi }{Q^2}\partial _zf \, ,
\end{align}
and  $\xi_{\mu }$ is the polarization, which can be decomposed as  
\begin{equation}
\xi_{\mu} = \xi_{\mu}^{\perp} + \frac{ q \cdot \xi}{Q^2} q_{\mu}\,.
\end{equation}

\section{Numeric convergence}
\label{sec:on_numerics}

\begin{figure}[!b]
	  \centering
    \includegraphics[scale=0.67]{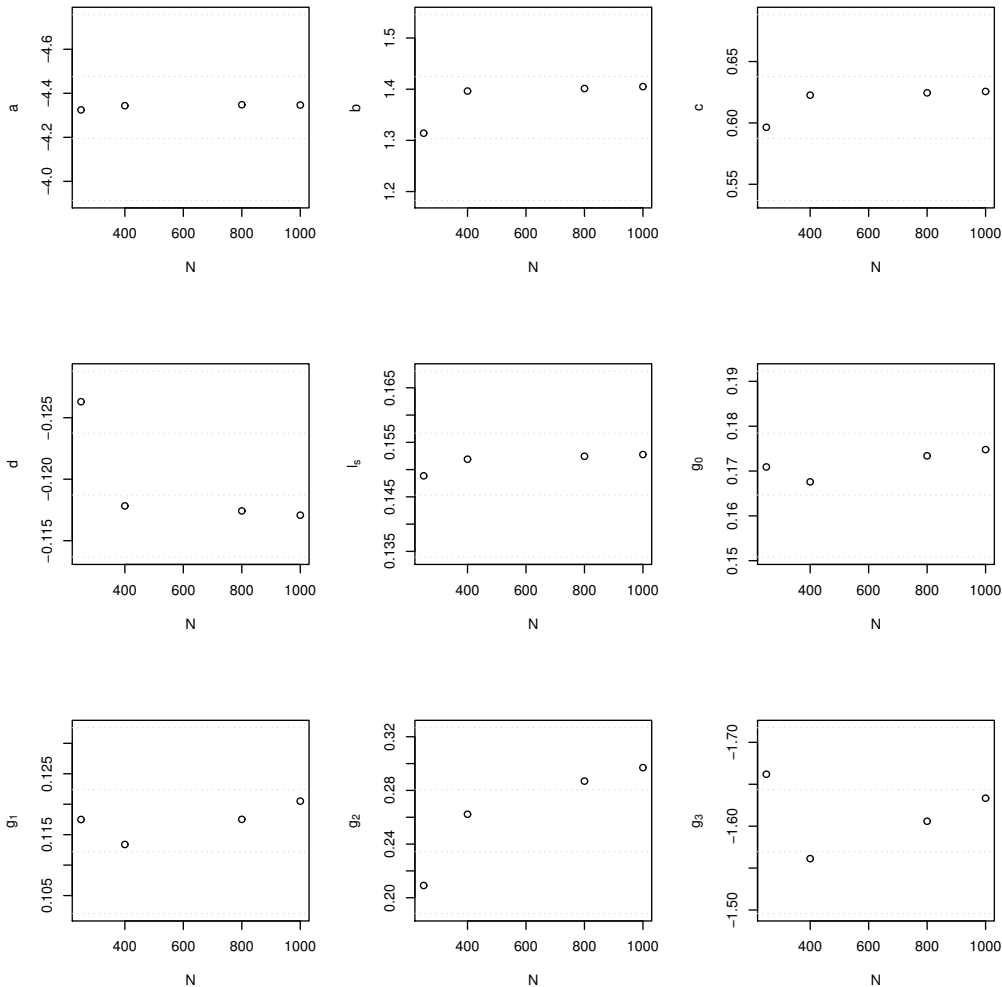}
		\caption{Evolution of best fit parameters with the number of interpolation points $N$.}
    \label{fig:parameters_evolution}
\end{figure}

For the specific model we consider in this paper, configurations for the potential with a very large dip close to $z=0$ appear for the 3rd and 4th trajectories, as can be seen in figure \ref{fig:spectrum}, requiring a careful analysis of the precision of the computation. As commented in the paper, we have used mainly a Chebyshev algorithm for solving the Schrödinger problem in which functions in the interval $[z_{min}, z_{max}]$ are discretized in $N$ points. We have done first our minimization procedure starting with $N=250$, and then  gradually increased it to $N=400$, $800$ and $1000$. At each one of these values of $N$, using as a starting point the best values obtained from the previous $N$, we run our minimization routine always obtaining a $\chi^2\sim1.7$. In figure \ref{fig:parameters_evolution} it is shown the evolution of the best fit parameters with $N$, in the paper we reported values for  
the $N=1000$ case.

% BIBLIOGRAPHY
% use BIBTEX if you want
\bibliographystyle{JHEP}
\bibliography{pomeron,HS,HQCD,integrability,glueballs,bootstrap,refs}

\providecommand{\href}[2]{#2}\begingroup\raggedright\begin{thebibliography}{10}

\bibitem{donnachie_pomeron_2002}
S.~Donnachie, H.~G. Dosch, O.~Nachtmann and P.~Landshoff, \emph{Pomeron physics
  and {QCD}}, {\emph{Camb.Monogr.Part.Phys.Nucl.Phys.Cosmol.} {\bf 19} (2002)
  1--347}.

\bibitem{donnachie_total_1992}
A.~Donnachie and P.~V. Landshoff, \emph{Total cross sections},
  \href{http://dx.doi.org/10.1016/0370-2693(92)90832-O}{\emph{Physics Letters
  B} {\bf 296} (Dec., 1992) 227--232}.

\bibitem{Aaron:2009aa}
{\scshape ZEUS, H1} collaboration, F.~D. Aaron et~al., \emph{{Combined
  Measurement and QCD Analysis of the Inclusive e+- p Scattering Cross Sections
  at HERA}}, \href{http://dx.doi.org/10.1007/JHEP01(2010)109}{\emph{JHEP} {\bf
  01} (2010) 109}, [\href{https://arxiv.org/abs/0911.0884}{{\tt 0911.0884}}].

\bibitem{donnachie_small_1998}
A.~Donnachie and P.~V. Landshoff, \emph{Small x: two pomerons!},
  \href{http://dx.doi.org/10.1016/S0370-2693(98)00899-5}{\emph{Physics Letters
  B} {\bf 437} (Oct., 1998) 408--416}.

\bibitem{donnachie_new_2001}
A.~Donnachie and P.~V. Landshoff, \emph{New data and the hard pomeron},
  \href{http://dx.doi.org/10.1016/S0370-2693(01)01048-6}{\emph{Physics Letters
  B} {\bf 518} (Oct., 2001) 63--71}.

\bibitem{donnachie_perturbative_2002}
A.~Donnachie and P.~V. Landshoff, \emph{Perturbative {QCD} and {Regge} theory:
  closing the circle},
  \href{http://dx.doi.org/10.1016/S0370-2693(02)01556-3}{\emph{Physics Letters
  B} {\bf 533} (May, 2002) 277--284}.

\bibitem{donnachie_elastic_2011}
A.~Donnachie and P.~V. Landshoff, \emph{Elastic {Scattering} at the {LHC}},
  {\emph{arXiv:1112.2485 [hep-ex, physics:hep-ph]} (Dec., 2011) }.

\bibitem{donnachie_pp_2013}
A.~Donnachie and P.~V. Landshoff, \emph{pp and total cross sections and elastic
  scattering},
  \href{http://dx.doi.org/10.1016/j.physletb.2013.10.068}{\emph{Physics Letters
  B} {\bf 727} (Dec., 2013) 500--505}.

\bibitem{Fadin:1975cb}
V.~S. Fadin, E.~A. Kuraev and L.~N. Lipatov, \emph{{On the Pomeranchuk
  Singularity in Asymptotically Free Theories}},
  \href{http://dx.doi.org/10.1016/0370-2693(75)90524-9}{\emph{Phys. Lett.} {\bf
  B60} (1975) 50--52}.

\bibitem{Kuraev:1977fs}
E.~A. Kuraev, L.~N. Lipatov and V.~S. Fadin, \emph{{The Pomeranchuk Singularity
  in Nonabelian Gauge Theories}}, {\emph{Sov. Phys. JETP} {\bf 45} (1977)
  199--204}.

\bibitem{Balitsky:1978ic}
I.~I. Balitsky and L.~N. Lipatov, \emph{{The Pomeranchuk Singularity in Quantum
  Chromodynamics}}, {\emph{Sov. J. Nucl. Phys.} {\bf 28} (1978) 822--829}.

\bibitem{kowalski_using_2010}
H.~Kowalski, L.~N. Lipatov, D.~A. Ross and G.~Watt, \emph{Using {HERA} {Data}
  to {Determine} the {Infrared} {Behaviour} of the {BFKL} {Amplitude}},
  \href{http://dx.doi.org/10.1140/epjc/s10052-010-1500-6}{\emph{The European
  Physical Journal C} {\bf 70} (Dec., 2010) 983--998}.

\bibitem{polchinski_hard_2002}
J.~Polchinski and M.~J. Strassler, \emph{Hard {Scattering} and {Gauge}/{String}
  {Duality}},
  \href{http://dx.doi.org/10.1103/PhysRevLett.88.031601}{\emph{Physical Review
  Letters} {\bf 88} (Jan., 2002) 031601}.

\bibitem{polchinski_deep_2003}
J.~Polchinski and M.~J. Strassler, \emph{Deep inelastic scattering and
  gauge/string duality},
  \href{http://dx.doi.org/10.1088/1126-6708/2003/05/012}{\emph{Journal of High
  Energy Physics} {\bf 2003} (May, 2003) 012}.

\bibitem{brower_pomeron_2007}
R.~C. Brower, J.~Polchinski, M.~J. Strassler and C.-I. Tan, \emph{The {Pomeron}
  and gauge/string duality},
  \href{http://dx.doi.org/10.1088/1126-6708/2007/12/005}{\emph{Journal of High
  Energy Physics} {\bf 2007} (Dec., 2007) 005}.

\bibitem{hatta_deep_2008}
Y.~Hatta, E.~Iancu and A.~H. Mueller, \emph{Deep inelastic scattering at strong
  coupling from gauge/string duality: the saturation line},
  \href{http://dx.doi.org/10.1088/1126-6708/2008/01/026}{\emph{Journal of High
  Energy Physics} {\bf 2008} (2008) 026}.

\bibitem{cornalba_saturation_2008}
L.~Cornalba and M.~S. Costa, \emph{Saturation in deep inelastic scattering from
  the {AdS}/{CFT} correspondence},
  \href{http://dx.doi.org/10.1103/PhysRevD.78.096010}{\emph{Physical Review D}
  {\bf 78} (Nov., 2008) 096010}.

\bibitem{pire_ads/qcd_2008}
B.~Pire, C.~Roiesnel, L.~Szymanowski and S.~Wallon, \emph{On {AdS}/{QCD}
  correspondence and the partonic picture of deep inelastic scattering},
  \href{http://dx.doi.org/10.1016/j.physletb.2008.10.026}{\emph{Physics Letters
  B} {\bf 670} (Dec., 2008) 84--90}.

\bibitem{albacete_dis_2008}
J.~L. Albacete, Y.~V. Kovchegov and A.~Taliotis, \emph{{DIS} on a large nucleus
  in {AdS}/{CFT}},
  \href{http://dx.doi.org/10.1088/1126-6708/2008/07/074}{\emph{Journal of High
  Energy Physics} {\bf 2008} (2008) 074}.

\bibitem{hatta_relating_2008}
Y.~Hatta, \emph{Relating e + e - annihilation to high energy scattering at weak
  and strong coupling},
  \href{http://dx.doi.org/10.1088/1126-6708/2008/11/057}{\emph{Journal of High
  Energy Physics} {\bf 2008} (2008) 057}.

\bibitem{levin_glauber-gribov_2009}
E.~Levin, J.~Miller, B.~Z. Kopeliovich and I.~Schmidt, \emph{Glauber-{Gribov}
  approach for {DIS} on nuclei in {N} = 4 {SYM}},
  \href{http://dx.doi.org/10.1088/1126-6708/2009/02/048}{\emph{Journal of High
  Energy Physics} {\bf 2009} (2009) 048}.

\bibitem{brower_saturation_2008}
R.~C. Brower, M.~Djuric and C.-I. Tan, \emph{Saturation and {Confinement}:
  {Analyticity}, {Unitarity} and {AdS}/{CFT} {Correspondence}},
  {\emph{arXiv:0812.1299 [hep-ph]} (Dec., 2008) }.

\bibitem{brower_elastic_2009}
R.~Brower, M.~Djuric and C.-I. Tan, \emph{Elastic and {Diffractive}
  {Scattering} after {AdS}/{CFT}}, {\emph{arXiv:0911.3463 [hep-ph]} (Nov.,
  2009) }.

\bibitem{gao_polarized_2009}
J.-H. Gao and B.-W. Xiao, \emph{Polarized deep inelastic and elastic scattering
  from gauge/string duality},
  \href{http://dx.doi.org/10.1103/PhysRevD.80.015025}{\emph{Physical Review D}
  {\bf 80} (July, 2009) 015025}.

\bibitem{hatta_polarized_2009}
Y.~Hatta, T.~Ueda and B.-W. Xiao, \emph{Polarized {DIS} in $\mathcal{N} = 4$
  {SYM}: where is spin at strong coupling?},
  \href{http://dx.doi.org/10.1088/1126-6708/2009/08/007}{\emph{Journal of High
  Energy Physics} {\bf 2009} (2009) 007}.

\bibitem{kovchegov_comparing_2009}
Y.~V. Kovchegov, Z.~Lu and A.~H. Rezaeian, \emph{Comparing {AdS}/{CFT}
  calculations to {HERA} ${F}_2$ data},
  \href{http://dx.doi.org/10.1103/PhysRevD.80.074023}{\emph{Physical Review D}
  {\bf 80} (Oct., 2009) 074023}.

\bibitem{avsar_shockwaves_2009}
E.~Avsar, E.~Iancu, L.~McLerran and D.~N. Triantafyllopoulos, \emph{Shockwaves
  and deep inelastic scattering within the gauge/gravity duality},
  \href{http://dx.doi.org/10.1088/1126-6708/2009/11/105}{\emph{Journal of High
  Energy Physics} {\bf 2009} (2009) 105}.

\bibitem{cornalba_deep_2010}
L.~Cornalba, M.~S. Costa and J.~Penedones, \emph{Deep inelastic scattering in
  conformal {QCD}},
  \href{http://dx.doi.org/10.1007/JHEP03(2010)133}{\emph{Journal of High Energy
  Physics} {\bf 2010} (Mar., 2010) 1--65}.

\bibitem{dominguez_particle_2010}
F.~Dominguez, \emph{Particle production in {DIS} off a shockwave in {AdS}},
  \href{http://dx.doi.org/10.1007/JHEP09(2010)007}{\emph{Journal of High Energy
  Physics} {\bf 2010} (Sept., 2010) 7}.

\bibitem{cornalba_ads_2010}
L.~Cornalba, M.~S. Costa and J.~Penedones, \emph{{AdS} {Black} {Disk} {Model}
  for {Small}-x {Deep} {Inelastic} {Scattering}},
  \href{http://dx.doi.org/10.1103/PhysRevLett.105.072003}{\emph{Physical Review
  Letters} {\bf 105} (Aug., 2010) 072003}.

\bibitem{betemps_diffractive_2010}
M.~A. Betemps, V.~P. Gon{\c c}alves and J.~T. de~Santana~Amaral,
  \emph{Diffractive deep inelastic scattering in an {AdS}/{CFT} inspired model:
  {A} phenomenological study},
  \href{http://dx.doi.org/10.1103/PhysRevD.81.094012}{\emph{Physical Review D}
  {\bf 81} (May, 2010) 094012}.

\bibitem{gao_polarized_2010}
J.-H. Gao and Z.-G. Mou, \emph{Polarized deep inelastic scattering off the
  neutron from gauge/string duality},
  \href{http://dx.doi.org/10.1103/PhysRevD.81.096006}{\emph{Physical Review D}
  {\bf 81} (May, 2010) 096006}.

\bibitem{kovchegov_$r$_2010}
Y.~V. Kovchegov, \emph{R-current dis on a shock wave: Beyond the eikonal
  approximation},
  \href{http://dx.doi.org/10.1103/PhysRevD.82.054011}{\emph{Physical Review D}
  {\bf 82} (Sept., 2010) 054011}.

\bibitem{levin_inelastic_2010}
E.~Levin and I.~Potashnikova, \emph{Inelastic processes in {DIS} and
  $\mathcal{N}=4$ {SYM}},
  \href{http://dx.doi.org/10.1007/JHEP08(2010)112}{\emph{Journal of High Energy
  Physics} {\bf 2010} (Aug., 2010) 112}.

\bibitem{domokos_pomeron_2009}
S.~K. Domokos, J.~A. Harvey and N.~Mann, \emph{Pomeron contribution to pp and
  $pp^{-}$ scattering in {AdS}/{QCD}},
  \href{http://dx.doi.org/10.1103/PhysRevD.80.126015}{\emph{Physical Review D}
  {\bf 80} (Dec., 2009) 126015}.

\bibitem{domokos_setting_2010}
S.~K. Domokos, J.~A. Harvey and N.~Mann, \emph{Setting the scale of the $pp$
  and $p\overline{p}$ total cross sections using {AdS}/{QCD}},
  \href{http://dx.doi.org/10.1103/PhysRevD.82.106007}{\emph{Physical Review D}
  {\bf 82} (Nov., 2010) 106007}.

\bibitem{brower_string-gauge_2010}
R.~C. Brower, M.~Djuric, I.~Sarcevic and C.-I. Tan, \emph{String-gauge dual
  description of deep inelastic scattering at small-x},
  \href{http://dx.doi.org/10.1007/JHEP11(2010)051}{\emph{Journal of High Energy
  Physics} {\bf 2010} (Nov., 2010) 1--26}.

\bibitem{costa_deeply_2012}
M.~S. Costa and M.~Djurić, \emph{Deeply virtual {Compton} scattering from
  gauge/gravity duality},
  \href{http://dx.doi.org/10.1103/PhysRevD.86.016009}{\emph{Physical Review D}
  {\bf 86} (July, 2012) 016009}.

\bibitem{Brower:2012mk}
R.~C. Brower, M.~Djuric and C.-I. Tan, \emph{{Diffractive Higgs Production by
  AdS Pomeron Fusion}},
  \href{http://dx.doi.org/10.1007/JHEP09(2012)097}{\emph{JHEP} {\bf 09} (2012)
  097}, [\href{https://arxiv.org/abs/1202.4953}{{\tt 1202.4953}}].

\bibitem{costa_vector_2013}
M.~S. Costa, M.~Djurić and N.~Evans, \emph{Vector meson production at low x
  from gauge/gravity duality},
  \href{http://dx.doi.org/10.1007/JHEP09(2013)084}{\emph{Journal of High Energy
  Physics} {\bf 2013} (Sept., 2013) 1--18}.

\bibitem{anderson_central_2014}
N.~Anderson, S.~K. Domokos, J.~A. Harvey and N.~Mann, \emph{Central production
  of $\eta$ and $\eta'$ via double {Pomeron} exchange in the {Sakai}-{Sugimoto}
  model}, \href{http://dx.doi.org/10.1103/PhysRevD.90.086010}{\emph{Physical
  Review D} {\bf 90} (Oct., 2014) 086010}.

\bibitem{Nally:2017nsp}
R.~Nally, T.~G. Raben and C.-I. Tan, \emph{{Inclusive Production Through
  AdS/CFT}},  \href{https://arxiv.org/abs/1702.05502}{{\tt 1702.05502}}.

\bibitem{gursoy_exploring_2008}
U.~Gürsoy and E.~Kiritsis, \emph{Exploring improved holographic theories for
  {QCD}: part {I}},
  \href{http://dx.doi.org/10.1088/1126-6708/2008/02/032}{\emph{Journal of High
  Energy Physics} {\bf 2008} (Feb., 2008) 032}.

\bibitem{gursoy_exploring_2008-1}
U.~Gürsoy, E.~Kiritsis and F.~Nitti, \emph{Exploring improved holographic
  theories for {QCD}: part {II}},
  \href{http://dx.doi.org/10.1088/1126-6708/2008/02/019}{\emph{Journal of High
  Energy Physics} {\bf 2008} (Feb., 2008) 019}.

\bibitem{gursoy_improved_2011}
U.~Gürsoy, E.~Kiritsis, L.~Mazzanti, G.~Michalogiorgakis and F.~Nitti,
  \emph{Improved {Holographic} {QCD}},
  \href{http://dx.doi.org/10.1007/978-3-642-04864-7_4}{\emph{arXiv:1006.5461
  [hep-lat, physics:hep-ph, physics:hep-th]} {\bf 828} (2011) 79--146}.

\bibitem{meyer_glueball_2005}
H.~B. Meyer, \emph{Glueball {Regge} {Trajectories}},
  {\emph{arXiv:hep-lat/0508002} (Aug., 2005) }.

\bibitem{meyer_glueball_2005-1}
H.~B. Meyer and M.~J. Teper, \emph{Glueball {Regge} trajectories and the
  pomeron: a lattice study},
  \href{http://dx.doi.org/10.1016/j.physletb.2004.11.036}{\emph{Physics Letters
  B} {\bf 605} (Jan., 2005) 344--354}.

\bibitem{Ballon-Bayona:2015wra}
A.~Ballon-Bayona, R.~Carcassés~Quevedo, M.~S. Costa and M.~Djurić,
  \emph{{Soft Pomeron in Holographic QCD}},
  \href{http://dx.doi.org/10.1103/PhysRevD.93.035005}{\emph{Phys. Rev.} {\bf
  D93} (2016) 035005}, [\href{https://arxiv.org/abs/1508.00008}{{\tt
  1508.00008}}].

\bibitem{BallonBayona:2007qr}
C.~A. Ballon~Bayona, H.~Boschi-Filho and N.~R.~F. Braga, \emph{{Deep inelastic
  scattering from gauge string duality in the soft wall model}},
  \href{http://dx.doi.org/10.1088/1126-6708/2008/03/064}{\emph{JHEP} {\bf 03}
  (2008) 064}, [\href{https://arxiv.org/abs/0711.0221}{{\tt 0711.0221}}].

\bibitem{BallonBayona:2008zi}
C.~A. Ballon~Bayona, H.~Boschi-Filho and N.~R.~F. Braga, \emph{{Deep inelastic
  scattering from gauge string duality in D3-D7 brane model}},
  \href{http://dx.doi.org/10.1088/1126-6708/2008/09/114}{\emph{JHEP} {\bf 09}
  (2008) 114}, [\href{https://arxiv.org/abs/0807.1917}{{\tt 0807.1917}}].

\bibitem{BallonBayona:2010ae}
C.~A. Ballon~Bayona, H.~Boschi-Filho, N.~R.~F. Braga and M.~A.~C. Torres,
  \emph{{Deep inelastic scattering for vector mesons in holographic D4-D8
  model}}, \href{http://dx.doi.org/10.1007/JHEP10(2010)055}{\emph{JHEP} {\bf
  10} (2010) 055}, [\href{https://arxiv.org/abs/1007.2448}{{\tt 1007.2448}}].

\bibitem{Koile:2013hba}
E.~Koile, S.~Macaluso and M.~Schvellinger, \emph{{Deep inelastic scattering
  structure functions of holographic spin-1 hadrons with $N_f \geq 1$}},
  \href{http://dx.doi.org/10.1007/JHEP01(2014)166}{\emph{JHEP} {\bf 01} (2014)
  166}, [\href{https://arxiv.org/abs/1311.2601}{{\tt 1311.2601}}].

\bibitem{Koile:2015qsa}
E.~Koile, N.~Kovensky and M.~Schvellinger, \emph{{Deep inelastic scattering
  cross sections from the gauge/string duality}},
  \href{http://dx.doi.org/10.1007/JHEP12(2015)009}{\emph{JHEP} {\bf 12} (2015)
  009}, [\href{https://arxiv.org/abs/1507.07942}{{\tt 1507.07942}}].

\bibitem{costa_conformal_2012}
M.~S. Costa, V.~Goncalves and J.~Penedones, \emph{Conformal {Regge} theory},
  \href{http://dx.doi.org/10.1007/JHEP12(2012)091}{\emph{Journal of High Energy
  Physics} {\bf 2012} (Dec., 2012) 1--50}.

\bibitem{Cornalba:2007fs}
L.~Cornalba, \emph{{Eikonal methods in AdS/CFT: Regge theory and multi-reggeon
  exchange}},  \href{https://arxiv.org/abs/0710.5480}{{\tt 0710.5480}}.

\bibitem{karch_linear_2006}
A.~Karch, E.~Katz, D.~T. Son and M.~A. Stephanov, \emph{Linear confinement and
  {AdS}/{QCD}},
  \href{http://dx.doi.org/10.1103/PhysRevD.74.015005}{\emph{Physical Review D}
  {\bf 74} (July, 2006) 015005}.

\bibitem{caron-huot_strings_2016}
S.~Caron-Huot, Z.~Komargodski, A.~Sever and A.~Zhiboedov, \emph{{Strings from
  Massive Higher Spins: The Asymptotic Uniqueness of the Veneziano Amplitude}},
   \href{https://arxiv.org/abs/1607.04253}{{\tt 1607.04253}}.

\end{thebibliography}\endgroup

\end{document}